\documentclass[final,3p,times,pdflatex]{elsarticle}
\usepackage{amsmath}
\usepackage{amssymb}
\usepackage{graphicx}
\usepackage{color}

\def\SOFTSUSY{{\tt SOFTSUSY}}
\def\code#1{{\tt #1}}

\journal{Computer Physics Communications}

\begin{document}

\begin{frontmatter}

\begin{flushright}
DAMTP-2014-42\\
IFIC/14-47
\end{flushright}

\title{Higher Order Corrections and Unification in
  the Minimal Supersymmetric Standard Model: {\tt SOFTSUSY3.5}}

\author[damtp]{B.C.~Allanach}
\author[dubna]{A.~Bednyakov}
\cortext[cor1]{Corresponding author}
\author[valencia]{R.~Ruiz~de~Austri\corref{cor1}}
\ead{rruiz@ific.uv.es}

\address[damtp]{DAMTP, CMS, University of Cambridge, Wilberforce road,
  Cambridge, CB3  0WA, United Kingdom}
\address[dubna]{Joint Institute for Nuclear Research, 141980, Dubna, Russia}
\address[valencia]{Instituto de Física Corpuscular, IFIC-UV/CSIC, E-46980
  Paterna, Spain}  

\begin{abstract}
  We explore the effects of three-loop minimal supersymmetric standard
  model   renormalisation group equation terms and some leading two-loop
  threshold corrections on gauge and Yukawa
  unification: each being one loop higher order than 
  current public spectrum calculators. We also explore the effect of the
  higher order terms (often 2-3 GeV) on the lightest
  CP even Higgs mass prediction.
  We illustrate our results in the constrained minimal supersymmetric standard
  model. Neglecting threshold corrections at the grand unified scale, the
  discrepancy between the unification scale
  $\alpha_s$ and 
  the other two unified gauge couplings changes by 0.1$\%$ due to the
  higher order 
  corrections and
  the difference between unification scale bottom-tau
  Yukawa couplings neglecting unification scale threshold
  corrections changes by up to 1$\%$. The difference between unification
  scale bottom and top Yukawa couplings changes by a few percent.
  Differences due to the higher order corrections also give an estimate of the
  size of theoretical uncertainties in the minimal supersymmetric standard
  model spectrum. We use these to 
  provide   estimates 
  of theoretical uncertainties in predictions of the dark matter relic density
  (which can be of order one due to its strong dependence on sparticle masses)
  and the LHC sparticle production cross-section (often around 30$\%$).
  The additional higher order corrections have been
  incorporated into {\tt SOFTSUSY}, and we provide details on how to
  compile and use the program. We also provide a summary of the
  approximations used in the higher order corrections. 
\end{abstract}

\end{frontmatter}

\section{Program Summary}
\noindent{\em Program title:} \SOFTSUSY{}\\
{\em Program obtainable   from:} {\tt http://softsusy.hepforge.org/}\\
{\em Distribution format:}\/ tar.gz\\
{\em Programming language:} {\tt C++}, {\tt fortran}\\
{\em Computer:}\/ Personal computer.\\
{\em Operating system:}\/ Tested on Linux 3.4.6\\
{\em Word size:}\/ 64 bits.\\
{\em External routines:}\/ At least {\tt GiNaC1.3.5}~\cite{ginac} and {\tt
  CLN1.3.1}~(both freely obtainable from \verb|www.ginac.de|).\\
{\em Typical running time:}\/ A minute per parameter point.\\
{\em Nature of problem:}\/ Calculating supersymmetric particle spectrum and
mixing parameters in the minimal supersymmetric standard
model. The solution to the renormalisation group equations must be consistent
with boundary conditions on supersymmetry breaking parameters, as
well as the weak-scale boundary condition on gauge 
couplings, Yukawa couplings and the Higgs potential parameters.\\
{\em Solution method:}\/ Nested iterative algorithm. \\
{\em Restrictions:} \SOFTSUSY~will provide a solution only in the
perturbative regime and it
assumes that all couplings of the model are real
(i.e.\ $CP-$conserving). If the parameter point under investigation is
non-physical for some reason (for example because the electroweak potential
does not have an acceptable minimum), \SOFTSUSY{} returns an error message.
The higher order corrections included are for the real $R-$parity conserving
minimal supersymmetric standard model (MSSM) only.\\
{\em CPC Classification:} 11.1 and 11.6.\\
{\em Does the new version supersede the previous version?:} Yes.\\
{\em Reasons for the new version:} Extension to include additional two
and three-loop terms.\\
{\em Summary of revisions:} 
All quantities in the minimal supersymmetric standard model are extended to
have three-loop renormalisation group equations (including 3-family mixing) in
the limit of real 
parameters and some leading two-loop threshold
corrections are incorporated to the third family Yukawa couplings and the
strong gauge coupling. 
\newpage

\section{Introduction}

The recent discovery of the Higgs boson~\cite{Aad:2012tfa,Chatrchyan:2012ufa}
and the measurement of its mass at around 
125-126 GeV~\cite{ATLAS-CONF-2013-014} solidify the important and well-known
question of how its mass is 
stabilised with respect to quantum corrections, which are expected to be of
order the largest fundamental mass scale divided by the 16$\pi^2$ loop factor.
In particular, the Planck mass at $\sim 10^{19}$ GeV 
is expected to be the largest such relevant mass scale. 
However, since a quantum field theoretic description of gravity does not
exist it is possible, if not expected, that our effective field theory
description breaks down and such huge corrections are absent
for some reason. 
In any case, mass scales associated with the string scale $\sim 10^{17}$ GeV
or the grand unified theory (GUT) scale $M_{GUT}\sim 10^{16}$ GeV reintroduce
the question of stability of 
the Higgs mass.  
Imposing softly-broken supersymmetry upon the Standard Model provides a
well-known answer to this question, and this approach has been pursued
with vigour in the literature and at various high energy colliders (see, for
example, Refs.~\cite{Aad:2013wta,Chatrchyan:2014lfa}), where 
the predicted Standard Model particles' supersymmetric partners are being
searched for. To date, no unambiguous direct collider signals of
supersymmetric particles have been found, and a significant portion of the
most interesting parameter space has been ruled out. 
In order to rule a parameter point out, one predicts sparticle masses using
a supersymmetric spectrum generator and then simulates various collisions,
comparing to data to see if the predicted signals are significantly excluded
or not (or conversely, to see if there is statistically significant evidence
for a signal). The accurate measurement of a Higgs boson
now has become an important constraint upon any supersymmetric model. 
In order for this constraint to be as useful and as accurate as possible, the
prediction of the MSSM Higgs masses needs to be as accurate as possible. 
With a current estimated theoretical uncertainty in its prediction of around 3
GeV for `normal' supersymmetric spectra (i.e.\ sparticles in the TeV range), 
a reduction in the theoretical uncertainty in the lightest CP even Higgs
mass\footnote{In a large part of parameter space the lightest CP even Higgs
  boson 
  behaves approximately like the Standard Model Higgs boson.}
prediction is welcome. 

There are currently several available sparticle generators: {\tt
  ISAJET}~\cite{Paige:2003mg}, 
{\tt 
  SOFTSUSY}~\cite{Allanach:2001kg,Allanach:2009bv,Allanach:2011de,Allanach:2013kza}, 
{\tt SPheno}~\cite{Porod:2003um},
{\tt SUSEFLAV}~\cite{Chowdhury:2011zr}, 
{\tt SUSPECT}~\cite{Djouadi:2002ze}
as well as tailor made generators {\tt FlexibleSUSY}~\cite{Athron:2014yba} and
{\tt SphenoMSSM}~\cite{Porod:2011nf} 
based on {\tt SARAH}~\cite{Staub:2011dp,Staub:2013tta}.
Even specialising to the minimal
supersymmetric standard model (MSSM) with real parameters, 
these programs have slightly different
approximations, resulting in numerical predictions that are not
identical~\cite{Allanach:2003jw,Allanach:2004rh,Belanger:2005jk}. 
Even when calculations have at the same headline order of approximation (for
example, two-loop renormalisation group equations (RGEs) and one-loop
threshold corrections at $M_Z$), 
legitimate differences can result from the fact that higher order corrections
contribute to the calculation implicitly in different ways. 
If we take the example of a one-loop threshold correction to, for example, 
the prediction of the stop mass, there are various
contributions from Standard 
Model and supersymmetric particles. If we consider an internal loop with
a gluino propagator, which mass do we use for the gluino? One achieves
numerically distinct results if one uses two of the obvious choices: the
pole mass or the modified dimensional reduction ($\overline{DR}$) running
mass. The difference between the two prescriptions is a two-loop threshold
effect, and so either choice is allowed if one is working only at one loop
threshold effect order (numerically this is equivalent to working to two-loop
order in the RGEs, which are enhanced by a large logarithm).
Such choices occur hundreds of times within the calculation, multiplying the
possibilities for numerical differences. Thus, the numerical differences
between the spectrum calculators gives a very rough estimate of the size of
theoretical uncertainties associated with the calculation. 

One obvious way to reduce such a theoretical uncertainty is to incorporate
higher order effects, pushing the associated theoretical uncertainty to yet
higher orders.
That is what we have done for the present paper: we have
picked some available higher order terms that are expected to affect the
predictions of the spectrum mass calculation, and included them in
{\tt SOFTSUSY3.5.1}. The previous version of the program, {\tt SOFTSUSY3.4.1}, 
contained two-loop RGEs and one-loop threshold corrections. 
The higher order terms that we have included in the present paper are: 
\begin{enumerate}
\item
Three-loop RGEs \cite{Jack:2004ch} to all soft and supersymmetry preserving
MSSM parameters,  
assuming that such parameters are real. Both the supersymmetric and
soft-breaking MSSM 
parameters  
contain the possibility of full three-family mixing. 
\item
The following two-loop threshold
corrections calculated in the (electroweak) gaugeless limit~\cite{Haestier:2005ja} of the MSSM\footnote{With the only exception being the top-quark mass, for which only strong interactions are taken into account.}
\begin{center}
\begin{tabular}{|c|c|}\hline
Name & Description \\ \hline
$\Delta m_t$ & $\mathcal O(\alpha_s^2)$~\cite{Bednyakov:2002sf,Bednyakov:2005kt} corrections to $m_t$.  \\
$\Delta \alpha_s$ & $\mathcal O(\alpha_s^2)$~\cite{Harlander:2005wm,Bauer:2008bj}, $\mathcal O(\alpha_s \alpha_{t,b})$~\cite{Bednyakov:2010ni} corrections to $\alpha_s$. \\
$\Delta m_b, m_\tau$ & $\mathcal O(\alpha_s^2)$~\cite{Bauer:2008bj,Bednyakov:2007vm}, $\mathcal O(\alpha_s \alpha_{t,b})$, $\mathcal O(\alpha_{t,b}^2)$, $\mathcal O(\alpha_t \alpha_b)$, 
$\mathcal O(\alpha_{t,b} \alpha_\tau)$~\cite{Bednyakov:2009wt} corrections to $m_b$. \\ &
$m_\tau$ includes $\mathcal O(\alpha_\tau^2)$  and $\mathcal O(\alpha_\tau \alpha_{t,b})$ \cite{Bednyakov:2009wt} corrections. 
 \\
\hline \end{tabular}
\end{center}
\end{enumerate}
For our phenomenological analysis, we take the superpotential of the MSSM to be:
\begin{equation}
W=\mu H_2 H_1 + Y_t Q_3 H_2 u_3 + Y_b Q_3 H_1 d_3 + Y_\tau L_3 H_1 e_3,
\end{equation}
where the chiral superfields of the MSSM have the 
following $G_{SM}=SU(3)_c\times SU(2)_L\times U(1)_Y$ quantum numbers
\begin{eqnarray}
L_i:&(1,2,-\frac{1}{2}),\quad {e}_i:&(1,1,1),\qquad\, Q_i:\,(3,2,\frac{1}{6}),\quad
{u}_i:\,({\bar 3},1,-\frac{2}{3}),\nonumber\\ {d}_i:&({\bar
  3},1,\frac{1}{3}),\quad 
H_1:&(1,2,-\frac{1}{2}),\quad  H_2:\,(1,2,\frac{1}{2}),
\label{fields}
\end{eqnarray}
$i \in \{ 1,2,3\}$ is a family index and we have neglected all Yukawa
couplings except those of the third family.
In the table above, $\alpha_s$ denotes the strong coupling constant, $m_t$ the
top mass and
$\alpha_{t,b,\tau}=Y_{t,b,\tau}^2 / (4 \pi)$. 
None of $\Delta m_t$, $\Delta \alpha_s$ or $\Delta m_b, m_\tau$ have been, to
the best of our knowledge, made available to the public in a supported
computer program before. 

We shall illustrate our results with two different assumptions about
supersymmetry breaking soft terms. The first is the
constrained minimal supersymmetric standard model (CMSSM), which makes a
simplifying 
assumption about the supersymmetry breaking terms: each soft supersymmetry
(SUSY) breaking
scalar mass is set to a common value $m_0$ at a high scale $M_{GUT}\sim
10^{16}$ GeV (defined
here to be the scale at which the electroweak gauge couplings unify), the
gaugino masses are set to a common value $M_{1/2}$ at $M_{GUT}$ and the 
SUSY breaking trilinear scalar couplings are all fixed to a value $A_0$ at
$M_{GUT}$. The other relevant input parameters are $\tan \beta$, the ratio of
the 
two Higgs vacuum expectation values, and the sign of a parameter $\mu$ that
appears in the Higgs potential (its magnitude is fixed by the empirically
measured central value of 
the $Z^0$ boson mass via the minimisation of 
the 
Higgs potential). 

GUTs make the gauge unification prediction 
\begin{equation}
\alpha_1(M_{GUT})=\alpha_2(M_{GUT})=\alpha_3(M_{GUT}),  \label{gaugeUn}
\end{equation}
where $\alpha_1$ is the hypercharge gauge coupling in the GUT normalisation
and
$\alpha_2$ is the $SU(2)_L$ gauge coupling. 
If one uses gauge couplings inferred from measurements near the electroweak
scale 
and 
evolves them with the Standard Model RGEs, Eq.~\ref{gaugeUn} is not satisfied:
the gauge couplings $\alpha_1$ and $\alpha_3$ meet at a very different
renormalisation scale than $\alpha_1$ and $\alpha_2$. However, if we instead
assume the MSSM and calculate the evolution of gauge couplings at one loop
order, the prediction Eq.~\ref{gaugeUn} agrees with data
well~\cite{Amaldi:1987fu}. Two-loop 
predictions spoil this good agreement~\cite{Shifman:1994py}, but discrepancies
between the 
equalities are small and easily explained by heavy particles present in
realistic GUTs which
are not far below the GUT scale, for example $\sim \mathcal
O(M_{GUT}/10)$~\cite{Hall:1995eq}.  
These particles (for example heavy coloured triplets that come from
spontaneous breaking of the GUT group) affect the running of the gauge
couplings between their mass and $M_{GUT}$. Since we do not know of their
existence in our effective MSSM field theory, and we do not know their mass,
these effects are not taken into account in a general MSSM gauge unification
calculation, allowing for some small apparent `GUT threshold corrections' 
instead.
In practice, we define $M_{GUT}$
to be the renormalisation scale $Q$ where $\alpha_1(Q)=\alpha_2(Q)$, allowing
$\alpha_3(M_{GUT})$ to 
differ by a small amount due to the unknown heavy GUT-scale threshold
corrections.  
Some GUTs such as $SU(5)$~\cite{su5,su5b} predict bottom-tau
Yukawa unification 
\begin{equation}
Y_b(M_{GUT})=Y_\tau(M_{GUT}), \label{bTau}
\end{equation}
because both
particles reside in the same multiplet. In larger GUTs such as
SO(10)~\cite{Carena:1994bv}, the top  
Yukawa coupling is unified with the other two:
\begin{equation}
Y_t(M_{GUT})=Y_b(M_{GUT})=Y_\tau(M_{GUT}). \label{tbtau}
\end{equation}
In a similar way to gauge unification, small GUT threshold corrections may
slightly spoil apparent Yukawa unification. We shall therefore bear in mind
that there 
may be small corrections to Eqs.~\ref{bTau}, \ref{tbtau}.

The effect of the three-loop RGEs upon the relative mass shifts 
of Snowmass (SPS) benchmark 
points~\cite{Allanach:2002nj} were presented and studied in 
Ref.~\cite{Jack:2004ch} 
without the inclusion of the two-loop threshold effects. 
The three-loop RGEs are enhanced by a large logarithm $\log M_{GUT}/M_Z$,
which effectively promotes them to the size of a two-loop threshold effect. 
Thus the additional higher order terms that were included in
Ref.~\cite{Jack:2004ch} were of the same size as other terms that were
missing in the calculation. Effects upon sparticle mass predictions of around
1-2$\%$ were typically found, although one point studied did have an 8$\%$
difference in the light stop mass at the SPS5 point. 
We go beyond this calculation by including the threshold effects, which are of
the same order as the three-loop effects.

Subsequently, one of us performed~\cite{Bednyakov:2010ni} a preliminary study
of the SPS 4 (CMSSM high
$\tan \beta$) benchmark
point~\cite{Allanach:2002nj}, with a modified version of {\tt SOFTSUSY} that
included both the three-loop RGEs and the two-loop thresholds that we consider
here. 1-2$\%$ mass shifts in the strongly interacting sparticles, a 3$\%$
correction to the higgsino mass and a 1$\%$ decrease in the lightest CP even
higgs boson mass was observed. 

Our purpose here is to provide a more extensive
study of the higher order effects as well as to present a public version of {\tt
  SOFTSUSY} that incorporates them, along with instructions on how to use it.
In particular, we shall study the effects on Yukawa and gauge unification,
whose accuracy is improved by the inclusion of the higher orders.
In the prediction of the sparticle mass spectrum, there are other, two-loop
direct mass 
threshold 
contributions of the same order as the ones that we have included. The mass
spectrum does not therefore increase in precision, but the shift in masses
observed 
is a good estimator for the size of the theoretical uncertainty induced by
such two-loop direct mass threshold contributions. We shall study these
uncertainties and the induced uncertainties on other observables.
Our Higgs mass prediction includes direct two-loop threshold effects of the
same order as the ones that we have included, and so its accuracy is improved
by the inclusion of higher orders. 
Using three-loop RGEs accompanied by the two-loop
Higgs mass corrections allows one to effectively re-sum three-loop
next-to-next-to-leading (NNL) logarithmic terms proportional to powers of the strong
gauge and third family Yukawa couplings (multiplied by $\frac{1}{(16 \pi^2)^3}
\log 
\Lambda_{SUSY}/M_Z$, where
$\Lambda_{SUSY}$ is a common SUSY-breaking scale). 
We go beyond the pioneering study Ref.~\cite{Bednyakov:2010ni} in the
following ways: in section~\ref{sec:results}, we show case the effect of the
higher order terms in a CMSSM  
focus-point~\cite{Chan:1997bi,Feng:1999mn,Feng:1999zg}, where a high
sensitivity to the precise value of the top Yukawa 
coupling leads to large order one uncertainties in the mass spectrum. 
We examine induced uncertainties in the predicted LHC production
cross-sections and the dark matter relic density.
We then present a detailed
breakdown in the case of a phenomenological MSSM point, where the effects of
the RGEs are 
small, allowing us to focus, to a reasonable approximation, purely on the size
of the threshold corrections. We then quantify the effects of the higher order
corrections in a CMSSM plane that is used by ATLAS to interpret their searches
for supersymmetric particles in order to get an idea of the size of the
corrections for generic points in parameter space that are not excluded by the
current experimental limits. 
We scan a high $\tan \beta$ CMSSM plane to illustrate the large effect that the
higher order terms can have upon the dark matter relic density prediction,
across parameter space. 
In the appendix, we give details on how to install and use the increased
accuracy mode in a publicly available version of {\tt SOFTSUSY}.

\section{Effects of Higher Order Terms
 \label{sec:results}} 
Here, we shall examine how the higher order terms change unification
predictions and the Higgs and sparticle spectrum. The accuracy of the
unification  calculation
is
improved with the additional terms, and we shall investigate how much they
affect the accuracy with which gauge and Yukawa unification is (or is not)
achieved. 

Two-loop threshold effects
are not included in the calculation of sparticle masses in any of the public
programs, indeed most of them have not been calculated yet (with the notable
exceptions of the squark and gluino
masses~\cite{Martin:2005ch,Martin:2006ub}). This means that  
we are missing some terms {\em of the same order}\/ as those that we include 
in our higher order corrections for these quantities (for example, two-loop
threshold corrections to squark and gluino masses). Thus we cannot claim to
have increased the accuracy of the sparticle mass predictions. 
Differences in sparticle masses due to the higher order corrections do give an estimate of
the size of the missing terms, however, and are therefore instructive. 
Their inclusion is also a necessary step for the future when the two and
three-loop sparticle mass threshold corrections are included. 
Nevertheless, by means of three-loop RGEs, NNL logarithms can be
fully re-summed in the available two-loop Higgs mass prediction by a convenient choice of renormalization scale $\mu \sim \Lambda_{SUSY}$. 
Due to this, contrary to superpartners, we directly probe three-loop contribution to
the Higgs mass. One can see that a two-loop correction to a quantity that
appears in a one-loop correction is of order three-loops, but can have a reasonably large effect on
the lightest CP even Higgs mass $m_{h^0}$. This is because in our scheme where
the renormalization scale is $\Lambda_{SUSY}$,
in the MSSM the
tree-level lightest CP even Higgs mass is suppressed  $m_{h^0}<M_Z$ and
the one-loop corrections 
(dominantly due to Yukawa interactions of the top squarks) are 
larger than one would naively expect, being numerically
of the same order of magnitude. Therefore, two-loop corrections to the
couplings appearing in the one-loop expression are numerically of
approximately the same 
size as the two-loop Feynman diagrammatic Higgs contributions that are
included in {\tt SOFTSUSY}. However, we cannot rule out that some of the
three-loop Feynman diagrammatic contributions are of the same size as these
effects.  

In the next subsection, we shall examine two parameter points in detail
before performing parameter scans to characterise more generally the size of
the higher order effects.
Throughout this paper, we fix the important Standard Model parameters as
follows at or near their central empirical values~\cite{PhysRevD.86.010001}:
the top  
quark pole mass $m_t=173.2$ 
GeV, the running bottom quark mass in the $\overline{MS}$ scheme
$m_b(m_b)=4.18$ GeV, the strong coupling in the $\overline{MS}$ scheme
$\alpha_s(M_Z)=0.1187$ where the $Z^0$ boson pole mass is fixed to
$M_Z=91.1876$ GeV, the Fermi decay constant of the muon $G_\mu=1.16637
\times 10^{-5}$ GeV$^{-2}$, the fine structure constant in the $\overline{MS}$
scheme $\alpha(M_Z)=1/127.916$ and the pole mass of the tau lepton
$m_\tau=1.77699$ GeV. 

\subsection{Dissection of the higher order effects at 
  benchmark points}
In order to dissect the various higher order points we first define
different approximations to the prediction as in Table~\ref{tab:approx}.
\begin{table}
\begin{center}
\begin{tabular}{|c|c|c|} \hline
Name & RGEs & Quantity \\ \hline
$Q$ & 2 & Standard {\tt SOFTSUSY3.4.1} calculation without higher orders \\
$Q_3$ & 3 & Only 3-loop RGEs are included, not the 2-loop threshold corrections\\
$Q_{\alpha_s}$ & 2 & Included 2-loop threshold corrections to $\alpha_s$\\
$Q_{m_t}$ & 2 & Only the 2-loop threshold corrections to $m_t$ are included\\
$Q_{m_b,m_\tau}$ &2 & Only the 2-loop threshold corrections to $m_{b}$ and
$m_{\tau}$ are included\\
$Q_\textrm{All}$ & 3& All higher order corrections\\
\hline\end{tabular}
\end{center}
\caption{Different approximations for the calculation of a quantity $Q$. The
  column headed `RGEs' labels the number of loops used in the MSSM RGEs.\label{tab:approx}}
\end{table}
In Table~\ref{tab:cmssm}, we show the effects of the higher order terms on a
CMSSM parameter point that is in the high $\tan \beta$ focus point region:
($m_0=7240$ GeV, $M_{1/2}=800$ GeV, $A_0=-6000$ GeV,
$\tan \beta=50$, $\mu>0$) with 
some rather attractive phenomenological properties: it has a high lightest CP
even Higgs mass of 
124.6 GeV, agreeing with the experimental
central value~\cite{ATLAS-CONF-2013-014}, once theoretical uncertainties (estimated to be around $\pm 3$
GeV~\cite{Allanach:2004rh}) 
have been taken into account.
It also has attractive dark matter properties: $\Omega_{CDM} h^2=0.122$ is
close to the central value inferred from cosmological observations. In 
addition, the gluino and squark masses are heavy enough so as to not be ruled
out 
by the LHC7/8 TeV data. Apart from these phenomenologically advantageous
properties, the point has a high value of $\tan \beta=50$, which may give the 
bottom and tau Yukawa corrections a higher impact than if $\tan \beta$ were
smaller. At higher values of $\tan \beta$, the bottom and tau Yukawa couplings
are roughly proportional to $\tan \beta \approx 1/\cos{\beta}$:
\begin{equation}
Y_b(M_Z) = \frac{\sqrt{2}m_b(M_Z)}{v \cos \beta}, \qquad
Y_\tau(M_Z) = \frac{\sqrt{2}m_\tau(M_Z)}{v \cos \beta}.
\end{equation}
\begin{table}
\begin{center}
\begin{tabular}{|c|ccccccc|}\hline
 \vphantom{\bigg[} & $m_h$   & $m_A$   & $m_{\tilde g}$   & $m_{\chi_1^0}$   & $m_{\chi_2^0}$   & $m_{\chi_3^0}$   & $m_{\chi_4^0}$   \\ \hline
 Q                & 124.6 & 2416 & 2015 &  363 &  702 & 1134 &1140\\
$\Delta_3$  &   +0.1 & +18.9 &  -1.5 &  -0.1 &  +0.3 & +37.2 &+36.6\\
$\Delta \alpha_s$   &  +1.1 & +411.2 & -49.5 &  +2.1 &  +7.6 & +729.4 &+724.7\\
$\Delta m_t$       & N/A & N/A & N/A & N/A & N/A & N/A & N/A  \\
$\Delta m_b, m_\tau$ &  +0.5 & +490.5 &  -0.7 &  +0.3 &  +2.3 & +188.7 &+186.4\\
$\Delta$ All       &  -2.2 & -347.5 & -49.7 & -18.4 & -291.7 & -718.8 &-421.6\\

\hline  \vphantom{\bigg[}  & $m_{{\tilde q}}$  & $m_{{\tilde t}_1}$  & $m_{{\tilde t}_2}$  & $m_{{\tilde b}_1}$  & $m_{{\tilde b}_2}$  & $m_{{\tilde \tau}_1}$  & $m_{{\tilde \tau}_2}$ \\ \hline
Q              &7322 & 4951 & 4245 & 4946 & 5542 & 6265 &5095\\
$\Delta_3$  &   -4.2 & -11.3 & -18.7 & -11.2 &  -7.2 &  -0.1 & -0.5\\
$\Delta \alpha_s$  &   -2.2 & -71.3 & -167.9 & -71.1 &  -0.3 &  +0.6 & +1.6\\
$\Delta m_t$      &  N/A & N/A & N/A & N/A & N/A & N/A & N/A \\
$\Delta m_b, m_\tau$&   -0.5 & +55.1 & -34.2 & +55.5 & +125.9 &  +3.8 &+10.2\\
$\Delta$ All      &   -6.6 &  +3.7 & +65.3 &  +3.9 & -46.3 &  +6.6 &+15.9\\

\hline  \vphantom{\bigg[}  & $m_{\chi_1}^\pm$  & $m_{\chi_2}^\pm$  & $g_3(M_{SUSY})$  & $Y_t(M_{SUSY})$  & $Y_b(M_{SUSY})$  & $Y_\tau(M_{SUSY})$  & $\mu(M_{SUSY})$     \\ \hline
 Q                    &  703 & 1141 & 1.001 & 0.811 & 0.639 & 0.512 & 1114\\
$\Delta_3$  &    +0 &  +37 & +0.000 & +0.001 & -0.000 & +0.000 &  +37\\
$\Delta \alpha_s$  &    +8 & +725 & -0.019 & +0.014 & -0.005 & +0.000 & +727\\
$\Delta m_t$      &  N/A & N/A & N/A & N/A & N/A & N/A& N/A\\
$\Delta m_b, m_\tau$&    +2 & +186 & -0.000 & +0.004 & -0.023 & -0.001 & +187\\
$\Delta$ All      &  -303 & -422 & -0.020 & -0.016 & +0.001 & -0.001 & -718\\

\hline  \vphantom{\bigg[}  & $\Omega_{CDM} h^2$  & $\sigma_{SUSY}^{TOT}$  & $M_{GUT}/10^{16}$  & $1/\alpha_{GUT}$  & & & \\ \hline Q                   &  53.2 &   1.2 & 1.678 & 25.686 & & & \\
$\Delta_3$   &  +7.2 &  +0.0 & +0.007 & +0.003 & & & \\
$\Delta \alpha_s$   & +292.8 &  +0.4 & -0.046 & +0.058 &  &  & \\
$\Delta m_t$   & N/A & N/A & N/A & N/A & & & \\
$\Delta m_b,m_\tau$   & +48.8 &  -0.0 & -0.009 & +0.011 &  &  & \\
$\Delta$ All   & -53.1 &  +0.4 & +0.087 & -0.067 &  &  & \\

\hline  \vphantom{\bigg[}  & $10\Delta (\alpha)$  & $\Delta Y_{b \tau}$  & $\Delta Y_{tb}$  & & & & \\ \hline Q                    & -0.005 & -0.130 & 0.158 &  & & & \\
$Q_{3}$  & -0.005 & -0.129 & +0.160 &  &  &  & \\
$Q_{\alpha_s}$  & -0.005 & -0.130 & +0.158 &  &  &  & \\
$Q_{m_t}$  & N/A & N/A & N/A &  &  &  &  \\
$Q_{m_b,m_\tau}$  & -0.005 & -0.143 & +0.184 &  &  &  & \\
$Q_\textrm{All}$  &  -0.013 & -0.120 & +0.140 &  &  &  & \\
\hline
\end{tabular}
\end{center}

\caption{\label{tab:cmssm} Differences due to the highest order terms
  (three-loop 
  RGEs for gauge and Yukawa couplings and two-loop threshold corrections to
  third family fermion masses and $g_3$) on various predicted quantities 
  in the focus point of the CMSSM for $m_0=7240$ GeV, $M_{1/2}=800$ GeV,
  $A_0=-6000$ GeV, $\tan \beta=50$, $\mu>0$.
  We display massive quantities in units of GeV.
  The first column details which higher order threshold corrections are
  included. For some quantity, we have defined $\Delta_3=Q_3-Q$,
  $\Delta \alpha_s=Q_{\alpha_s}-Q$, $\Delta m_t=Q_{m_t}-Q$, 
  $\Delta m_b, 
  m_\tau=Q_{m_b m_\tau}-Q$ and $\Delta$ All$=Q_\textrm{All}-Q$ (see
  Table~\ref{tab:approx}). 
  `N/A' means that electroweak
  symmetry was not broken, and so reliable results cannot be reported. 
  $m_{\tilde q}$ refers to the average mass of the squarks of the first two
  families. 
The column headed
  `$\sigma_{SUSY}^{TOT}$' shows the total cross-section in fb for the
  production of   gluinos and squarks at a 14 TeV LHC.}
\end{table}
We split the various higher order corrections up in
the table: the `base' calculation is taken to be {\tt
  SOFTSUSY3.4.1}, which does not contain the higher order corrections. 
By comparing the entries for $\Delta_3$ with entries in other rows, we see that
three-loop RGE corrections on their own tend to induce smaller changes to the
quantities listed in the table than the threshold corrections, which tend to
be an order of magnitude larger in absolute value
than the two-loop threshold corrections
We can see that the two-loop threshold corrections to the strong coupling
$\alpha_s$ and top-quark mass $m_t$ are the most important ones.  
It is worth mentioning that both two-loop contributions to
$\alpha_s$ and $m_t$ (see., e.g., \cite{Pierce:1996zz,Bednyakov:2010ni}) 
tend to decrease the corresponding running MSSM parameters
at the matching scale.  
We see that the row $\Delta m_t$, that includes two-loop threshold corrections
to $m_t$ coming from strong SUSY QCD interactions only contains `N/A' entries,
indicating that the calculation failed in this approximation because
electroweak symmetry was not broken successfully, as we now explain.

Minimising the MSSM Higgs potential with respect to the electrically neutral
components of the Higgs vacuum expectation values, one obtains the well-known
tree-level result for the Higgs mass parameter $\mu$ in the modified
dimensional reduction 
scheme $(\overline{DR})$
\begin{equation}
\mu^2 = 
 \frac{\tan 2\beta}{2} \left[m_{{H}_2}^2\tan \beta
- m_{{H}_1}^2  \cot \beta \right] - \frac{M_Z^2}{2}. \label{higgsMin}
\end{equation}
In order to reduce\footnote{This prescription at least ensures that the
  dominant terms do not involve large logarithms.} missing higher order
corrections, 
all quantities in Eq.~\ref{higgsMin} are understood to be evaluated at a
$\overline{DR}$ renormalisation scale $Q=M_{SUSY}$, where $M_{SUSY}$ is the
geometric mean of the two stop masses.  
$\tan \beta=\langle H_2^0 \rangle / \langle H_1^0 \rangle$ is the ratio of the
two MSSM Higgs vacuum expectation values
 and $m_{{H}_{1,2}}$ are the soft SUSY breaking $\overline{DR}$ mass terms of
 the Higgs doublets.
If $m_{H_1}^2$ and $m_{H_2}^2$ and $\tan 2 \beta$ are such that $\mu^2>0$
results from Eq.~\ref{higgsMin}, the
model point may break electroweak symmetry successfully. On the other hand, if
$\mu^2 \le 0$, electroweak symmetry is {\em not}\/ broken successfully and the
model point is ruled out.
At the focus point,
the predicted value of $\mu$ derived from electroweak symmetry breaking is
known to 
depend extremely sensitively upon the precise value of the top Yukawa
coupling~\cite{Allanach:2000ii}. 
The parameter point in Table~\ref{tab:cmssm}
appears to agree with the experimental result on the Higgs
mass (which, in the CMSSM at high masses, acts to a good approximation with
identical couplings to the Standard Model Higgs)
according to the {\tt SOFTSUSY3.4.1}
calculation, bearing the $\pm 3$ GeV theoretical uncertainty in mind. However,
one would discard the point based on the predicted value 
of 53.2 for $\Omega_{CDM} h^2$, which disagrees with the cosmologically inferred
value by hundreds of sigma. On the other hand, including all of the high order
corrections (`$\Delta$ All'), we see that the Higgs mass prediction lowers
somewhat, and 
the dark matter 
relic density is predicted to be the cosmologically acceptable value of 0.122
once all of the higher order corrections are included.
Here, we use {\tt
  micrOMEGAs3.3.13}~\cite{Belanger:2001fz,Belanger:2004yn,Belanger:2013oya} to 
predict the relic density of 
lightest neutralinos, identified to be our dark matter candidate. 
Fits to cosmological data constrain the relic density of dark matter to be
$\Omega_{CDM} h^2=0.1198 \pm 0.0026$ from Planck data~\cite{Ade:2013zuv} (We
allow a $\pm 0.02$ error in the prediction coming from higher order
annihilation diagrams~\cite{Baro:2009na}).
Most of the CMSSM parameter space that is allowed by current sparticle
searches predicts a relic density that is far too high compared with
observations. However, there are isolated regions of parameter space that, for
one reason or another, have an enhanced annihilation mechanism where the dark
matter annihilates efficiently. In the focus point, the enhanced annihilation
comes from the fact that the dark matter candidate (the lightest neutralino)
has a significant higgsino component: small but real
values of $\mu(M_{SUSY})$ lead to a higgsino-dominated lightest neutralino dark
matter candidate, which annihilates efficiently into $WW$, $ZZ$, $Zh$ or $t
\bar t$~\cite{Feng:2000gh}, reducing the dark matter relic density to an
acceptable value. It also co-annihilates with the lightest chargino and the
second lightest neutralino. 
The MSSM Lagrangian contains the neutralino mass matrix as
$-\frac{1}{2}
{\tilde\psi^0}{}^T{\cal M}_{\tilde\psi^0}\tilde\psi^0$ + h.c., where
$\tilde\psi^0 =$ $(-i\tilde b,$ $-i\tilde w^3,$ $\tilde h_1,$ $\tilde
h_2)^T$ and, at tree level,
\begin{equation}
{\cal M}_{\tilde\psi^0} \ =\ \left(\begin{array}{cccc} M_1 & 0 &
-M_Zc_\beta s_W & M_Zs_\beta s_W \\ 0 & M_2 & M_Zc_\beta c_W &
-M_Zs_\beta c_W \\ -M_Zc_\beta s_W & M_Zc_\beta c_W & 0 & -\mu \\
M_Zs_\beta s_W & -M_Zs_\beta c_W & -\mu & 0
\end{array} \right),\label{mchi0}
\end{equation}
where $M_1$ and $M_2$ are the bino and wino SUSY breaking soft mass
parameters, respectively.
We use $s$ and $c$ for sine and cosine, so that
$s_\beta\equiv\sin\beta,\ c_{\beta}\equiv\cos\beta$ and $s_W (c_W)$ is
the sine (cosine) of the weak mixing angle. When $\mu \sim {\mathcal
  O}(\text{min}(M_1,\ M_2))$, 
the lightest mass eigenstate thus picks up a significant higgsino component, 
which has enhanced annihilation into the channels mentioned above. At high
$m_0$, other annihilation channels involving a $t-$channel scalar, are 
suppressed due to the high scalar mass. 
As each higher order correction is added, the value of $Y_t$ changes slightly,
changing the value of $\mu$ eventually predicted by Eq.~\ref{higgsMin}. 
Moving down the rows in Table~\ref{tab:cmssm} through each successive
approximation, we move from an approximation 
where $\mu(M_{SUSY}) > M_1$ and $\mu(M_{SUSY}) >
M_2$ to a situation where $\mu^2<0$ (the $\Delta m_t$ row, where `N/A' is
listed) to the approximation where all of our higher order corrections are
included, and $\mu(M_{SUSY})$ is of a similar magnitude to $M_1$ and
we have mixed 
higgsino-bino dark matter with an observationally acceptable predicted value.
In the literature,  a value of 2-3 GeV is often quoted as the spectrum
calculators' theoretical uncertainty on the prediction of $m_h$. We see that
this is borne out in our CMSSM model, where the higher order corrections give
a 2.2 GeV shift in the prediction.

Two-loop threshold corrections to the strong gauge coupling have a significant
effect upon some of the sparticle masses: particularly $m_{\chi_3^0}$ and
$m_{\chi_4^0}$. 
$m_{\chi_3^0}$ and
$m_{\chi_4^0}$ are controlled to leading order by $\mu$ (when $\mu>M_1, M_2)$,
which in turn is affected sensitively by the value of $m_{H_2}^2(M_{SUSY})$,
as Eq.~\ref{higgsMin} shows. $m_{H_2}^2(Q)$ runs very quickly with
renormalisation scale $Q$~\cite{Martin:1993zk}:
 \begin{eqnarray}
16 \pi^2\frac{\partial m_{H_2}^2}{d \ln Q} &=& 
6 \left[ (m_{H_2}^2 + m_{{\tilde Q}_3}^2 +
  m_{{\tilde u}_3}^2 + A_t^2)  Y_t^2  \right] - 6 g_2^2 M_2^2 - \frac{6}{5}
g_1^2 M_1^2 + \frac{3}{5}
g_1^2 \left(m_{H_2}^2-m_{H_1}^2 + \right. \nonumber \\
&&\left. \mbox{Tr}[m_{\tilde Q}^2 - m_{\tilde L}^2 - 2 
m_{\tilde u}^2 + m_{\tilde d}^2 + m_{\tilde e}^2] \right), \label{rgesA}
\end{eqnarray}
to one-loop order, and it is strongly affected by the value of the 
top Yukawa coupling $Y_t$.
The precise value of the top Yukawa coupling is affected by the strong
threshold corrections to the top quark mass
through~\cite{Pierce:1996zz}
\begin{equation}
Y_t(M_Z) = \frac{\sqrt{2} m_t(M_Z)}{v \sin \beta}, \qquad 
m_t = m_t(M_Z) + \Sigma_t(M_Z),
\end{equation}
where $\Sigma_t(M_Z)$ represents the MSSM top quark mass threshold corrections
(which include the strong threshold corrections).
The two-loop threshold corrections to the bottom and tau Yukawa couplings 
have an effect, particularly on $m_{\chi_3^0}$ and $m_{\chi_4^0}$ through
$\mu$ in a similar way to the effect due to $\Delta m_t$, as explained above.
In addition, the masses of heavy higgs bosons are shifted by about ten percent.
It is due to the fact that the running mass of CP-odd neutral higgs $m_A$ 
is given at tree-level by the well-known relation 
\begin{eqnarray}
	m_A^2 = \frac{1}{\cos 2 \beta} \left( m_{H_2}^2 - m_{H_1}^2 \right) - M_Z^2,
	\label{eq:mA}
\end{eqnarray}
into which both $m_{H_1}^2$ and $m_{H_2}^2$ enter on equal footing. 
Since the running of $m_{H_1}^2$ depends on the bottom Yukawa coupling \cite{Martin:1993zk},
the reduction of the corresponding boundary condition at $M_Z$ leads via RGE
to comparatively larger values of $m_{H_1}^2$ at $Q=M_{SUSY}$, which, in turn, 
increase\footnote{Both $\cos 2\beta$ in \eqref{eq:mA} and $\tan 2\beta$ in \eqref{higgsMin} are negative for $\tan\beta>1$.} the value of $m_A^2$. 
Although one cannot discern it from our table, 
the effect due to $\Delta m_t$ is apparently in the opposite direction and
tends to compensate 
that of $\Delta m_b$ \cite{Bednyakov:2010ni}.  
The lightest sbottom mass undergoes a 1$\%$ relative change,
whereas the other masses 
are less affected by two loop contribution to $m_b$. 

The smaller $\Delta \alpha_s$ changes in the gluino and average squark masses
lead to associated 
changes in the squark and gluino production cross-sections. 
For the point in Table~\ref{tab:cmssm}, SUSY particle production is dominated
by the production of two gluinos and 
their subsequent decay, the squarks being too heavy to be produced with any
appreciable cross-section. We calculate the next-to-leading order total QCD
cross section for production at a 14 TeV LHC with {\tt
  PROSPINO}~\cite{Beenakker:1996ed,Beenakker:1996ch}. We see that there is
a large 30$\%$ increase due to the higher order effects modifying $\alpha_s$,
which in turn changes the gluino mass. We emphasise again that the changes in
the 
spectrum that we see as a result of the higher order corrections are
indicative of the size of theoretical uncertainties in 
each mass prediction, but that the results with the higher order corrections
are only as accurate as those without them. 
However, our results on Yukawa and gauge couplings and their
unification {\em are} more accurate. 

For brevity, we have defined 
\begin{equation}
\Delta
(\alpha)=\alpha_3(M_{GUT})-\alpha_1(M_{GUT}),\qquad
\Delta Y_{b\tau}=Y_b(M_{GUT})-Y_\tau(M_{GUT}),\qquad
\Delta Y_{tb}=Y_t(M_{GUT})-Y_b(M_{GUT}),
\end{equation}
where $\alpha_{GUT}$ in the table refers to
$\alpha_1(M_{GUT})=\alpha_2(M_{GUT})$. 
Table~\ref{tab:cmssm} shows that the threshold corrections to $\alpha_s$ and
three-loop RGEs change $M_{GUT}$ and the unified gauge coupling $\alpha_{GUT}$
slightly. The discrepancy between $\alpha_{3}$ and $\alpha_1$ is generally
small, and is not affected much by the higher order corrections. The discrepancy
between the third family Yukawa couplings relatively decreases by some
20-30$\%$ once all of 
our higher order corrections are taken into account. 

\begin{table}
\begin{center}
\begin{tabular}{|c|ccccccc|}\hline
  & $m_h$  & $m_{\tilde g}$ & $m_{{\tilde q}}$ & $m_{\chi_1^0}$  & $m_{\chi_2^0}$ & $m_{\chi_3^0}$ & $m_{\chi_4}^0$ \\ \hline
$Q$               &  125.8 & 1087 & 1041 &  790 & 2428 & 2499 &2550\\
$\Delta_3$ &   +0.1 &  +0.0 &  +0.0 &  -0.0 &  -0.0 &  -0.0 & -0.0\\
$\Delta \alpha_s$  &   +1.6 &  -3.3 &  -1.6 &  +0.0 &  -0.2 &  -0.4 & -0.3\\
$\Delta m_t$      &  -2.4 &  +0.1 &  +0.0 &  -0.0 &  +0.3 &  +0.6 & +0.4\\
$\Delta m_b, m_\tau$&  +0.8 &  +0.0 &  +0.0 &  +0.0 &  -0.1 &  -0.2 & -0.1\\
$\Delta$ All      &  -1.4 &  -3.3 &  -1.6 &  -0.0 &  +0.1 &  +0.3 & +0.2\\

\hline& $m_{{\tilde t}_L}$  & $m_{{\tilde t}_R}$ &$m_{{\tilde b}_L}$&$m_{{\tilde b}_R}$&$m_{{\tilde \tau}_L}$&$m_{{\tilde \tau}_R}$&$m_{\chi_1}^\pm$ \\ \hline
$Q$             & 2616 & 2322 & 2497 & 2535 & 2523 & 2499 &2429\\
$\Delta_3$ &  -0.0 &  -0.3 &  -0.1 &  +0.0 &  +0.0 &  +0.0 & -0.0\\
$\Delta \alpha_s$  &  -1.1 &  -4.5 &  -2.1 &  -0.9 &  -0.0 &  +0.0 & -0.2\\
$\Delta m_t$      &  +0.3 &  +5.2 &  +1.7 &  +0.2 &  +0.0 &  -0.0 & +0.3\\
$\Delta m_b, m_\tau$&  -0.1 &  -1.6 &  -0.4 &  -0.1 &  +0.0 &  -0.0 & -0.1\\
$\Delta$ All      &  -0.7 &  +1.9 &  +0.2 &  -0.8 &  +0.0 &  -0.0 & +0.1\\

\hline      &   $g_3(M_{SUSY})$ & $Y_t(M_{SUSY})$ &  $Y_b(M_{SUSY})$ & $Y_\tau(M_{SUSY})$  & $\mu(M_{SUSY})$    & $\Omega_{CDM} h^2$ & $\sigma_{SUSY}^{TOT}$\\ \hline
 $Q$                   & 1.036 & 0.826 & 0.133 & 0.100 & 2500 &  0.12 & 1691.5\\
$\Delta_3$     &  +0.000 & +0.001 & +0.000 & +0.000 &   +0 & +0.00 &  +0.0\\
$\Delta \alpha_s$  & -0.014 & +0.012 & +0.001 & -0.000 &   +0 & +0.00 & +32.5\\
$\Delta m_t$      & +0.000 & -0.016 & +0.000 & +0.000 &   +0 & +0.00 &  +0.0\\
$\Delta m_b, m_\tau$& +0.000 & +0.005 & -0.003 & +0.000 &   +0 & +0.00 &  +0.0\\
$\Delta$ All      & -0.013 & -0.008 & -0.002 & +0.000 &   +0 & +0.00 &+32.5\\
\hline
\end{tabular}
\end{center}

\caption{\label{tab:pmssm}  Differences due to the highest order terms
  (three-loop 
  RGEs for gauge and Yukawa couplings and two-loop threshold corrections to
  third family fermion masses and $\alpha_s$) 
  in a modified phenomenological MSSM point 1.6~\cite{AbdusSalam:2011fc} (see
  text for the changes). The first
  column details which higher order threshold corrections are 
  included.   We display massive quantities in units of GeV.
  For some quantity $Q$, $\Delta_3=Q_3-Q$,
  $\Delta \alpha_s=Q_{\alpha_s}-Q$, $\Delta m_t=Q_{m_t}-Q$, 
  $\Delta m_b, 
  m_\tau=Q_{m_b m_\tau}-Q$ and $\Delta$ All$=Q_\textrm{All}-Q$ (see
  Table~\ref{tab:approx}). 
    $m_{\tilde q}$ refers to the average mass of the squarks of the first two
  families. 
  The rows marked with a $\Delta$ show the {\em change}\ with respect to
  the {\tt SOFTSUSY3.4.1} prediction. The column headed
  `$\sigma_{SUSY}^{TOT}$' shows the total cross-section in fb for the
  production of   gluinos and squarks at a 14 TeV LHC.}
\end{table}
We now wish to decouple the "pure" three-loop RGE effects from the two-loop threshold
effects as far as possible while still giving a valid prediction for a point
in MSSM parameter space. This can be achieved by studying the spectrum at a
point in pMSSM parameter space, where supersymmetry
breaking 
boundary conditions are imposed already at the SUSY breaking scale, defined to 
be $\sqrt{m_{{\tilde t}_1} m_{{\tilde t}_2}}$. We use the point pMSSM1.6 from
Ref.~\cite{AbdusSalam:2011fc}, which is defined to have all scalar trilinear
couplings set to 0, the tree-level first two generation squark masses, 
and tree-level gluino masses all set to 960 GeV and a tree-level bino
mass of 800 GeV. The tree-level wino mass $\mu$ and all other tree-level
squark and slepton masses are fixed to 2500 
GeV. We change the point however, the pseudo-scalar Higgs
pole mass is fixed to $M_{A^0}=1580$ GeV: this provides efficient dark matter
annihilation of two neutralinos (whose mass is approximately $M_{A^0}/2$).
We also change the stop mixing parameter, setting it to be $A_t=-5$ TeV, which
puts the prediction of the lightest CP even higgs mass to be near the
experimental central value.
The aim of using the pMSSM is to reduce the effects of the RGEs in order
to study the threshold contributions more cleanly. However, we cannot
eliminate RGE effects completely because there is still running between 
$M_Z$ and $M_{SUSY} \sim 2.5$ TeV. However, these running effects are small,
being of order 
$1/(16 \pi^2) \log M_{SUSY}/M_Z$, i.e.\ not enhanced by large logarithms,
unlike the CMSSM case above. 
We see that the lightest CP-even Higgs mass prediction decreases by 1.4 GeV,
mainly because of the  corrections to the top mass. 
Three-loop RGEs appear unnecessary here, only inducing relative changes of
order $10^{-4}$ in the predictions.
The high order strong
corrections to $\alpha_s$ reduce
squark and gluino masses at the per-mille level, which will only have a very
small effect on collider signatures. 
We see that the RGE corrections do induce a small additional change: but it is
at the per-mille or smaller level for this point.
The sparticle mass predictions only change by a
very small amount, which is contrary to the case of the focus-point CMSSM
(which is 
admittedly very sensitive to small changes in the top Yukawa coupling) that is
shown in Table~\ref{tab:cmssm}. 
The effect on the total 14 TeV LHC gluino/squark production cross-section is
only  around the $1\%$ level or so.
We conclude that in this point of the pMSSM, the higher order effects are not
needed for collider studies except for those involving the lightest CP-even
Higgs. We find it likely, where no input mass parameters are lighter
than 700 GeV that this conclusion will hold more generally for the
pMSSM because of sparticle decoupling in the corrections. 
However, to be sure of this conclusion, one should calculate the spectrum
at any point in question including
the higher order effects in order check. 
We shall show next that even for more
generic CMSSM points, there are typically relative changes in the spectra of
2-3$\%$. 

\subsection{CMSSM parameter scans}

\begin{figure}
\unitlength=1in
\begin{center}
\begin{picture}(7,2)
\put(-0.5,2){\includegraphics[angle=270,width=0.48\textwidth]{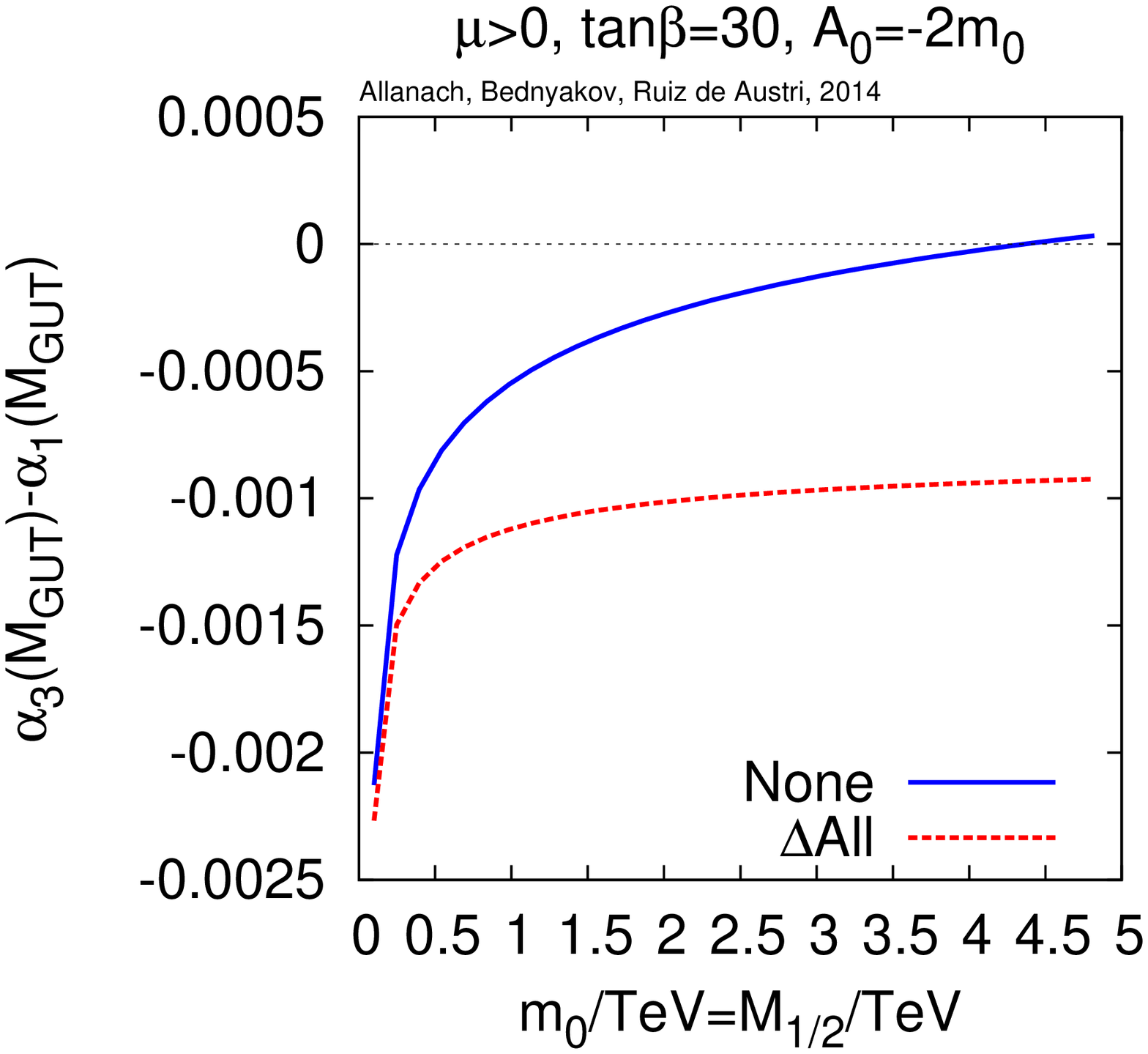}}
\put(1.83,2){\includegraphics[angle=270,width=0.48\textwidth]{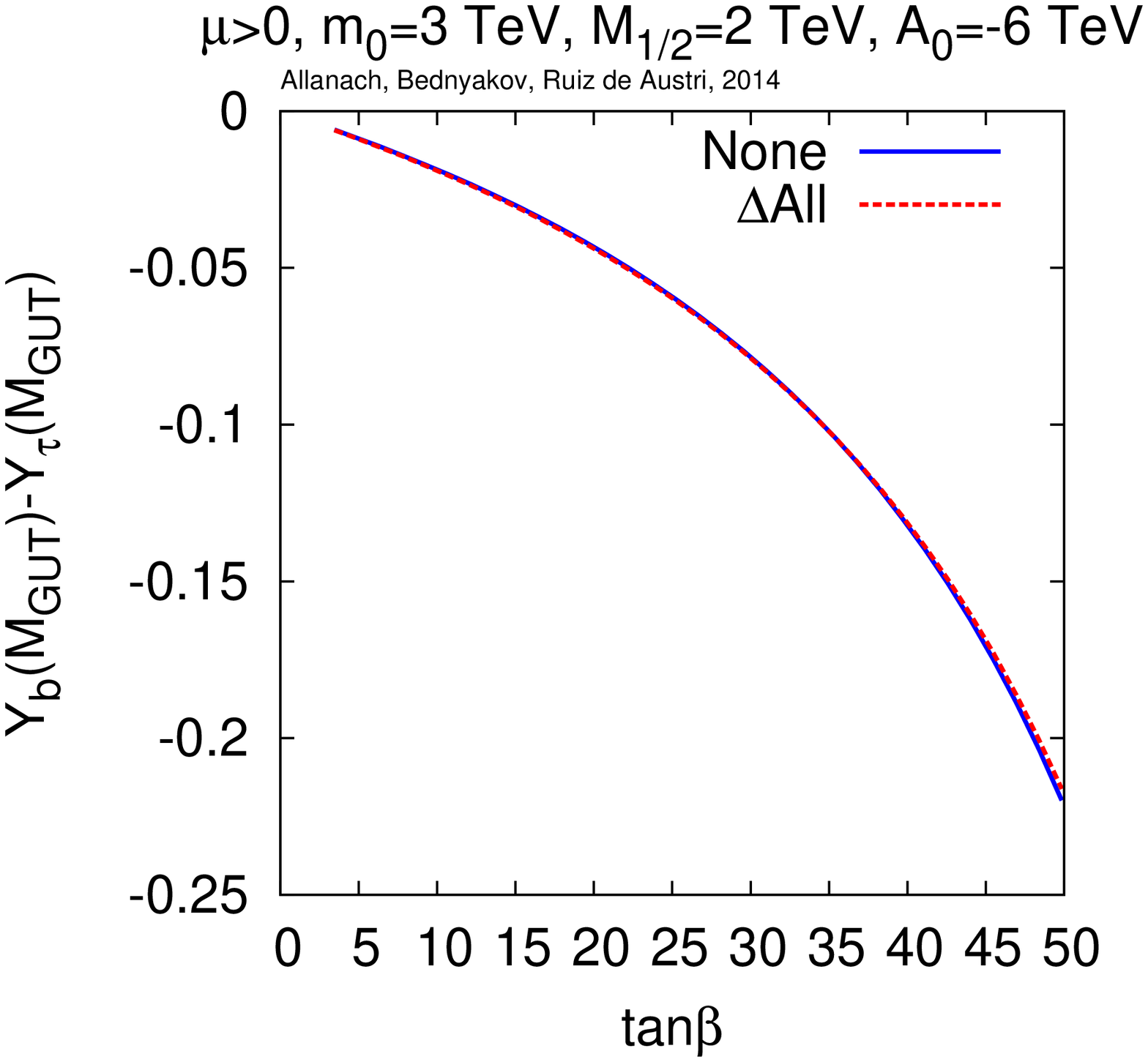}}
\put(4.16,2){\includegraphics[angle=270,width=0.48\textwidth]{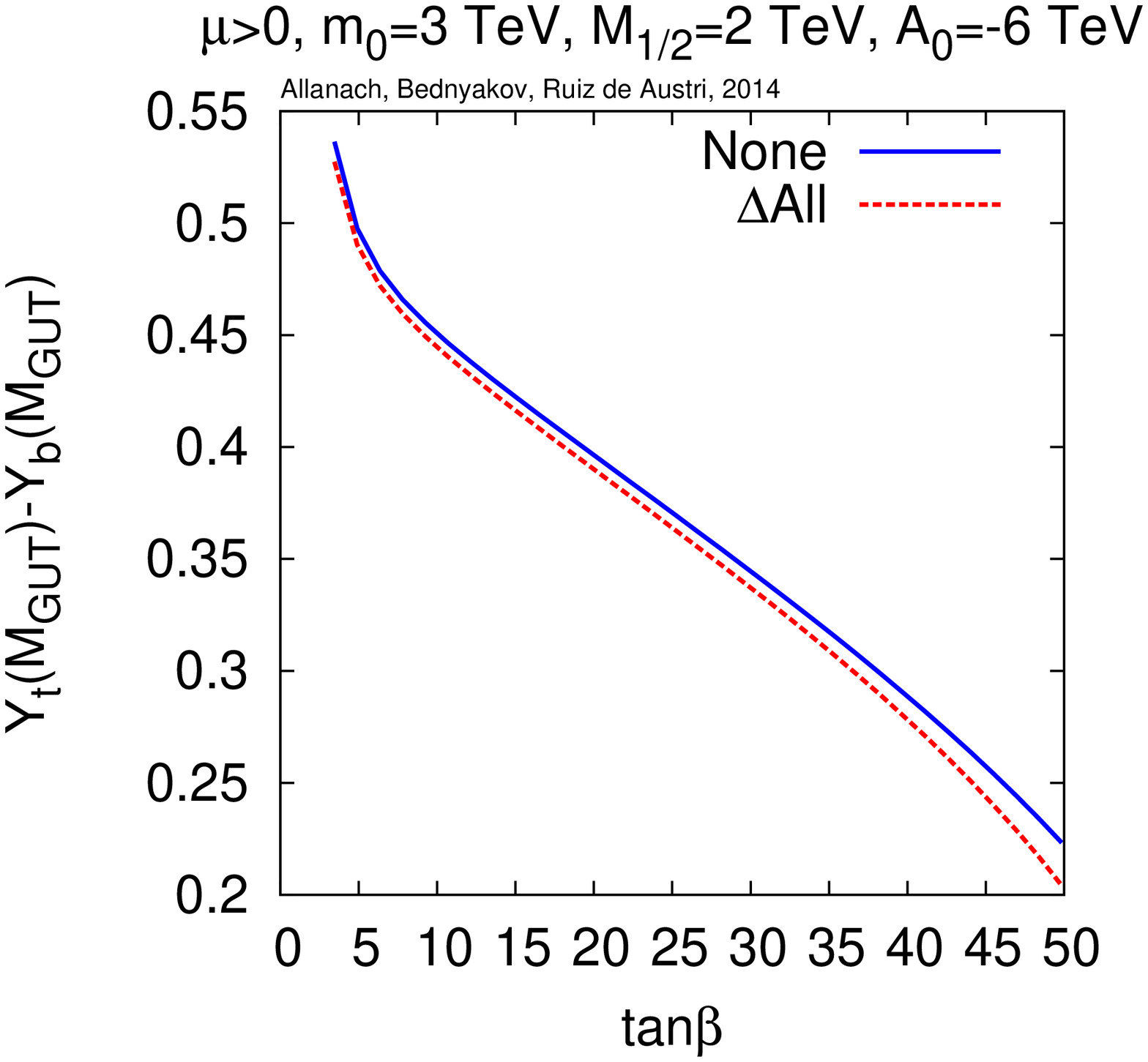}}
\put(0,1.9){(a)}
\put(2.33,1.9){(b)}
\put(4.66,1.9){(c)}
\end{picture}
\end{center}
\caption{\label{fig:uni}
Effects of higher order terms on unification in the CMSSM. The legend defines
whether the higher order terms are included (`$\Delta$All') or not (`None').}
\end{figure}
We now perform scans over CMSSM parameter space in order to examine the
effects of the higher order terms on unification. 
We shall often consider scans through CMSSM parameter space
around a 
$m_0-M_{1/2}$ parameter plane with $\mu>0$, $\tan \beta=30$ and $A_0=-2m_0$,
which was 
recently used by ATLAS to place bounds upon the CMSSM from various 8
TeV,  LHC `jets plus missing energy
searches'~\cite{Aad:2014wea} with 20 fb$^{-1}$ of
integrated luminosity.  
We see in Fig.~\ref{fig:uni}a that exact gauge unification occurs, for
$A=-2m_0$, $\mu>0$ and $\tan \beta=30$, at around $m_0=M_{1/2}\approx 4.5$ TeV
in the CMSSM at the usual {\tt SOFTSUSY3.4.1} accuracy. However, including the
higher order terms, $\alpha_3(M_{GUT})$ 
as predicted by data is around 0.001 times smaller than the other gauge
couplings. Such a discrepancy may be explained within a more detailed GUT model
via GUT threshold effects, but the precise value of
$\alpha_3(M_{GUT})-\alpha_1(M_{GUT})$ is important for constraining these. 
We see from Fig.~\ref{fig:uni}b that bottom-tau Yukawa unification is only
possible at low values of $\tan \beta$. There, however, the Higgs mass
prediction is too low compared with recent measurements. The higher order
corrections change the prediction of the $Y_b-Y_\tau$ Yukawa coupling
difference (to be  
acquired through GUT threshold effects) very little. 
Top-bottom Yukawa GUT unification is not possible for $\tan \beta < 40$,
as is evident in Fig.~\ref{fig:uni}c 
where $Y_t(M_{GUT})-Y_b(M_{GUT})$ is too large to be
explained by small loop effects. If the high-scale thresholds are instead well
below the GUT 
scale, perhaps a large enough correction may be possible at high $\tan \beta
> 40$. The higher order
effects make a large difference of several percent at high $\tan \beta$, and 
would significantly change the constraints upon these thresholds. 

\begin{figure}
\unitlength=1in
\begin{center}
\begin{picture}(6,7.8)
  \put(-0.75,2.8){\includegraphics[angle=270,width=0.7\textwidth]{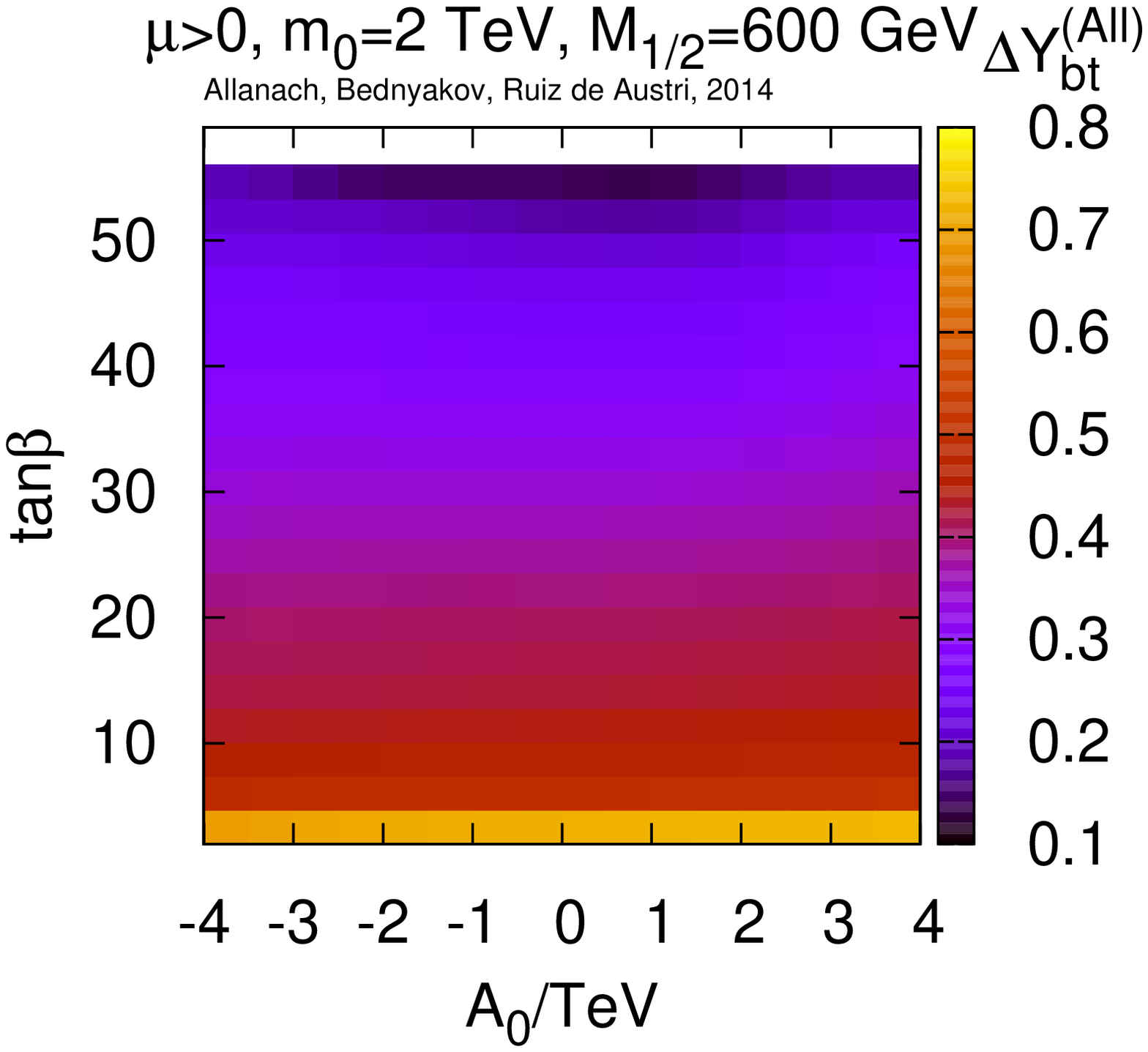}}
  \put(-0.75,5.5){\includegraphics[angle=270,width=0.7\textwidth]{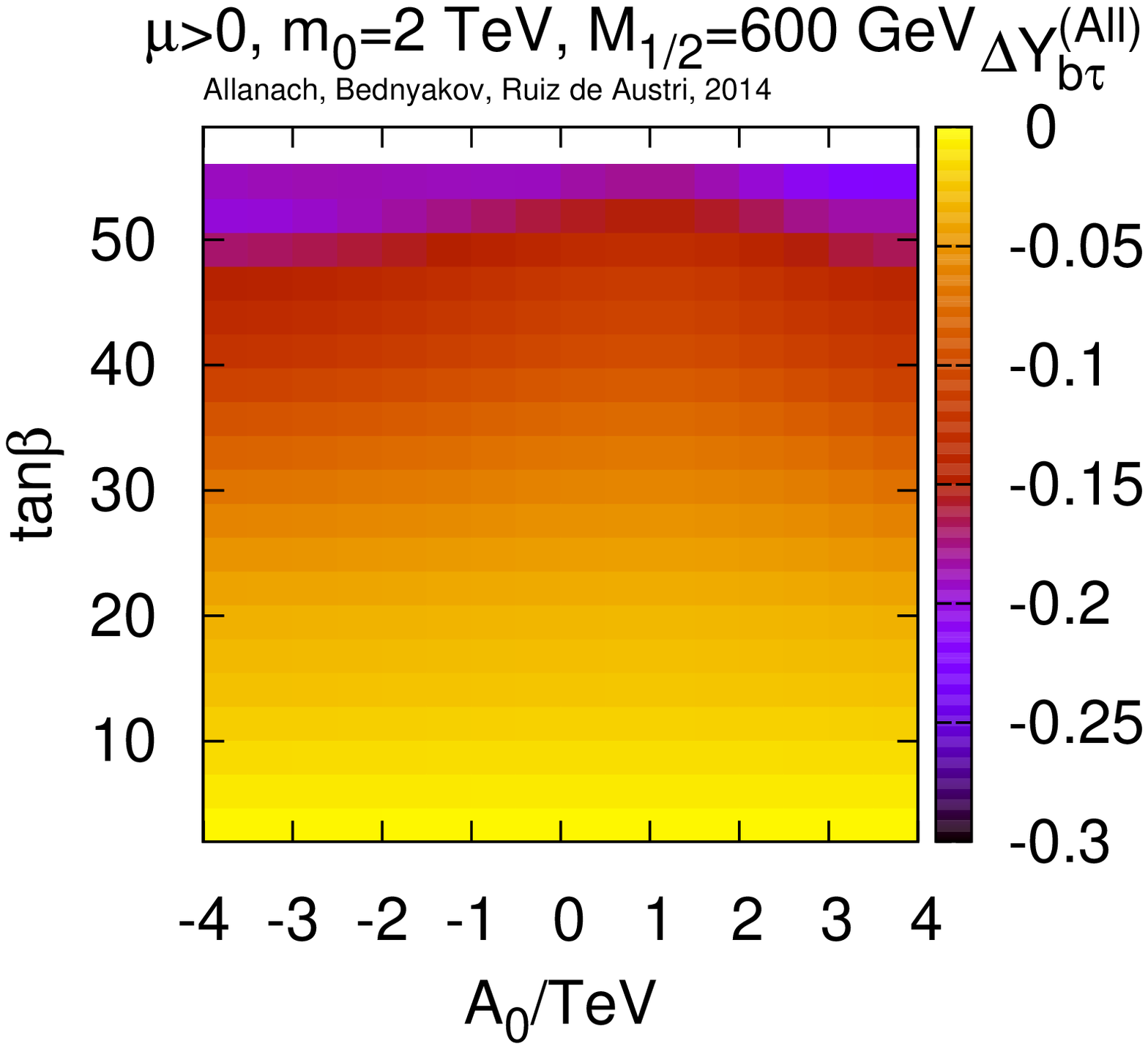}}
  \put(2.4,2.8){\includegraphics[angle=270,width=0.7\textwidth]{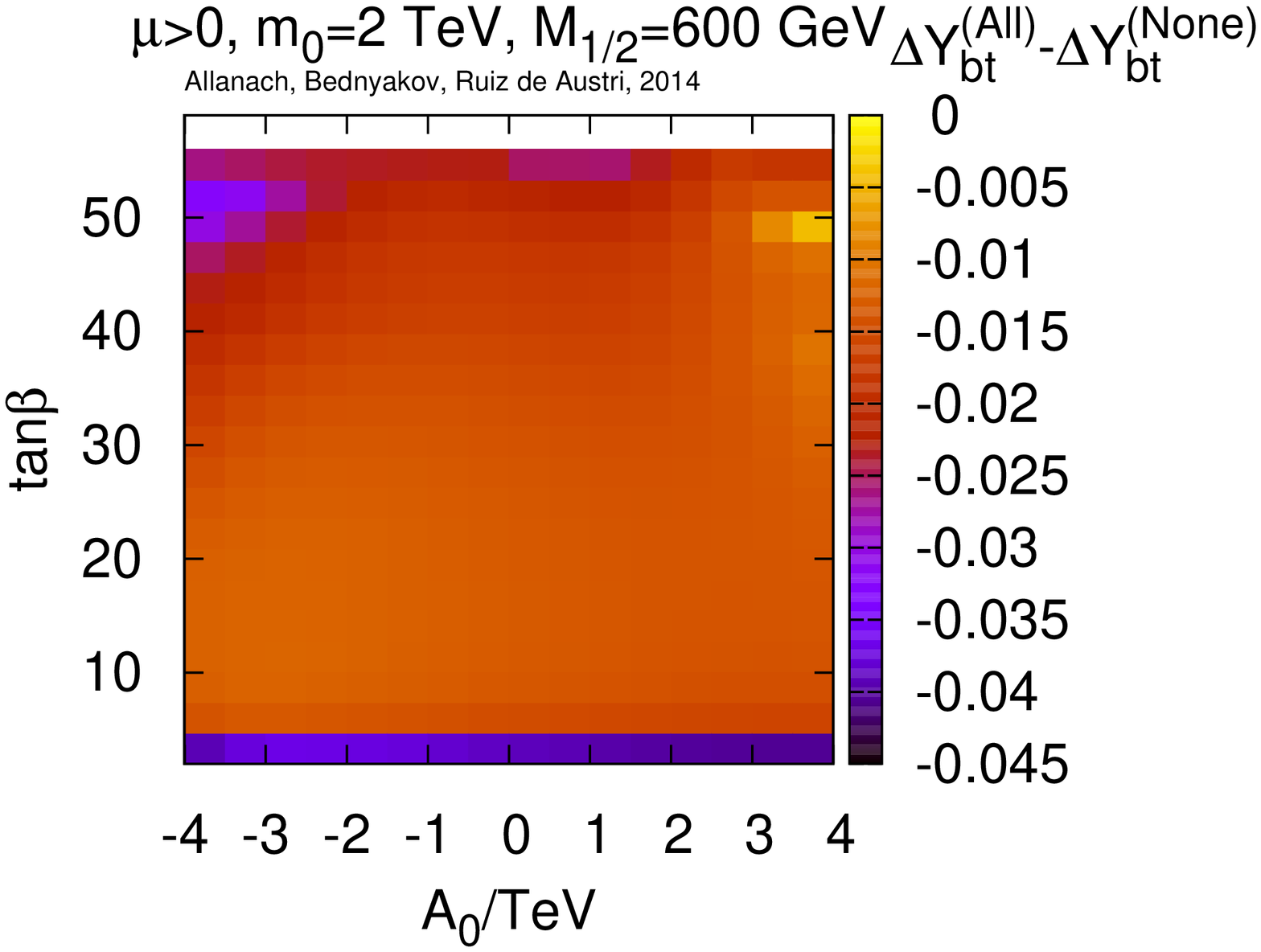}}
\put(2.4,5.5){\includegraphics[angle=270,width=0.7\textwidth]{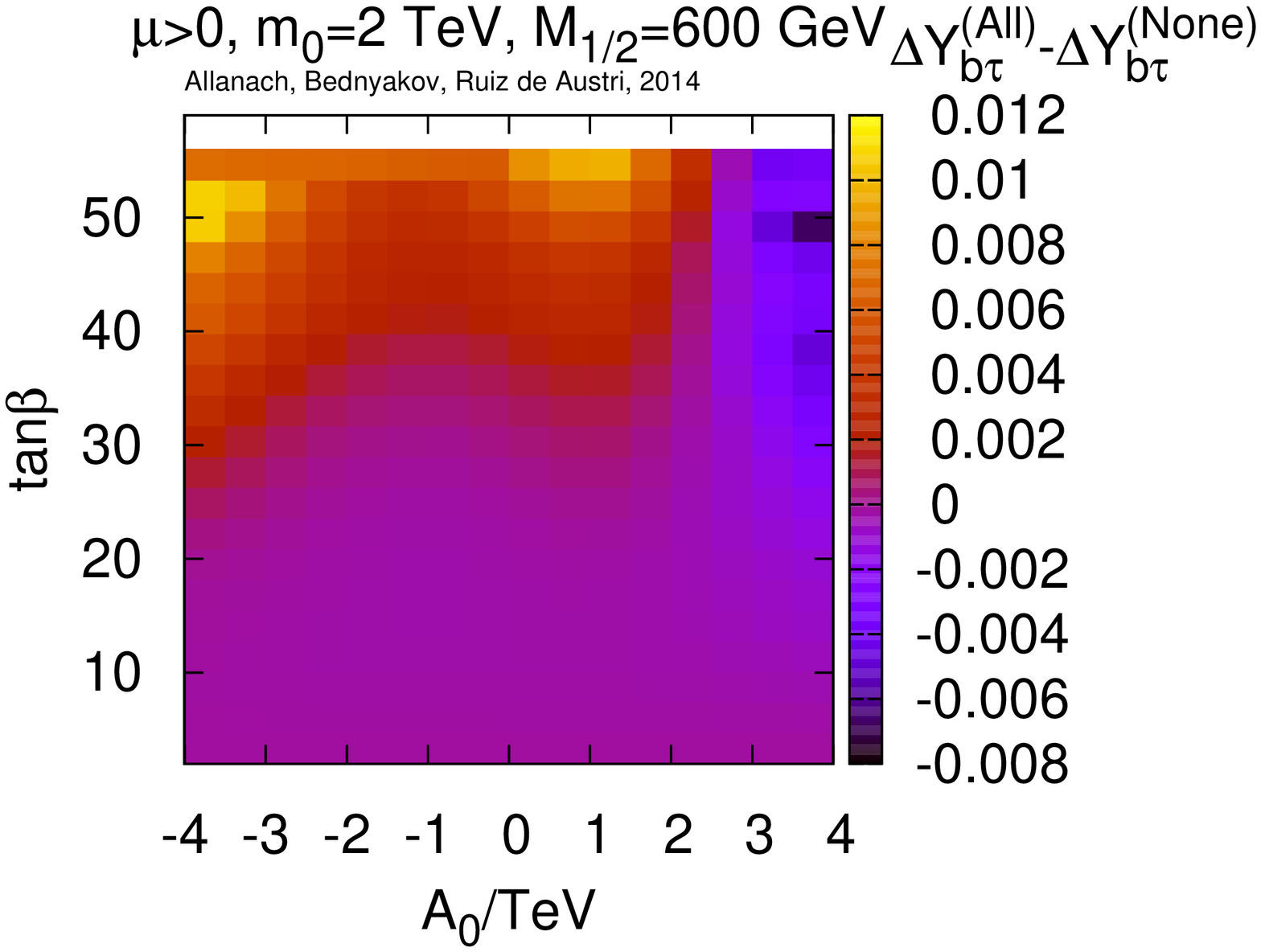}}
  \put(-0.75,8.2){\includegraphics[angle=270,width=0.7\textwidth]{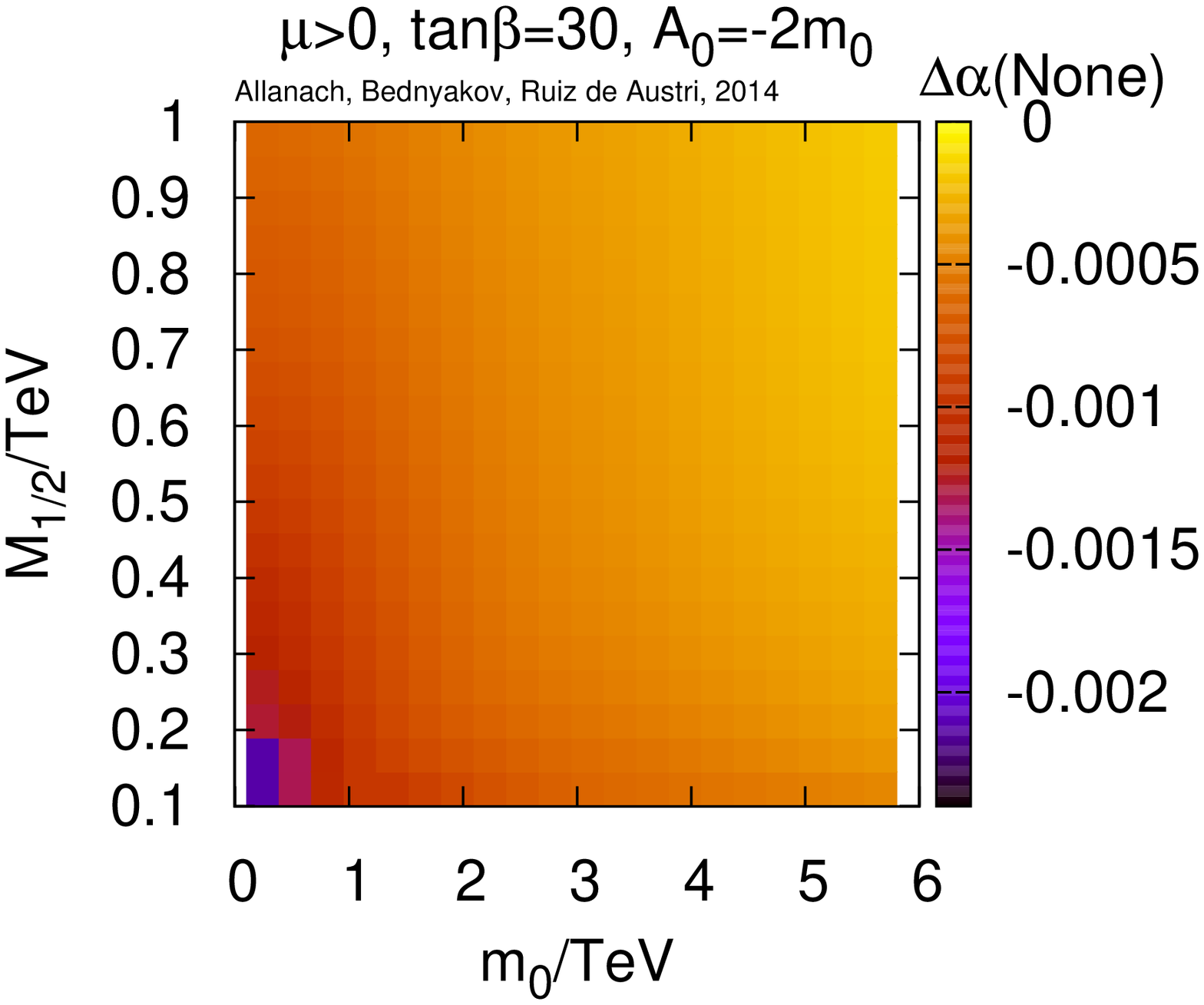}}
  \put(0.08,7.64){\includegraphics[angle=270,width=0.45\textwidth]{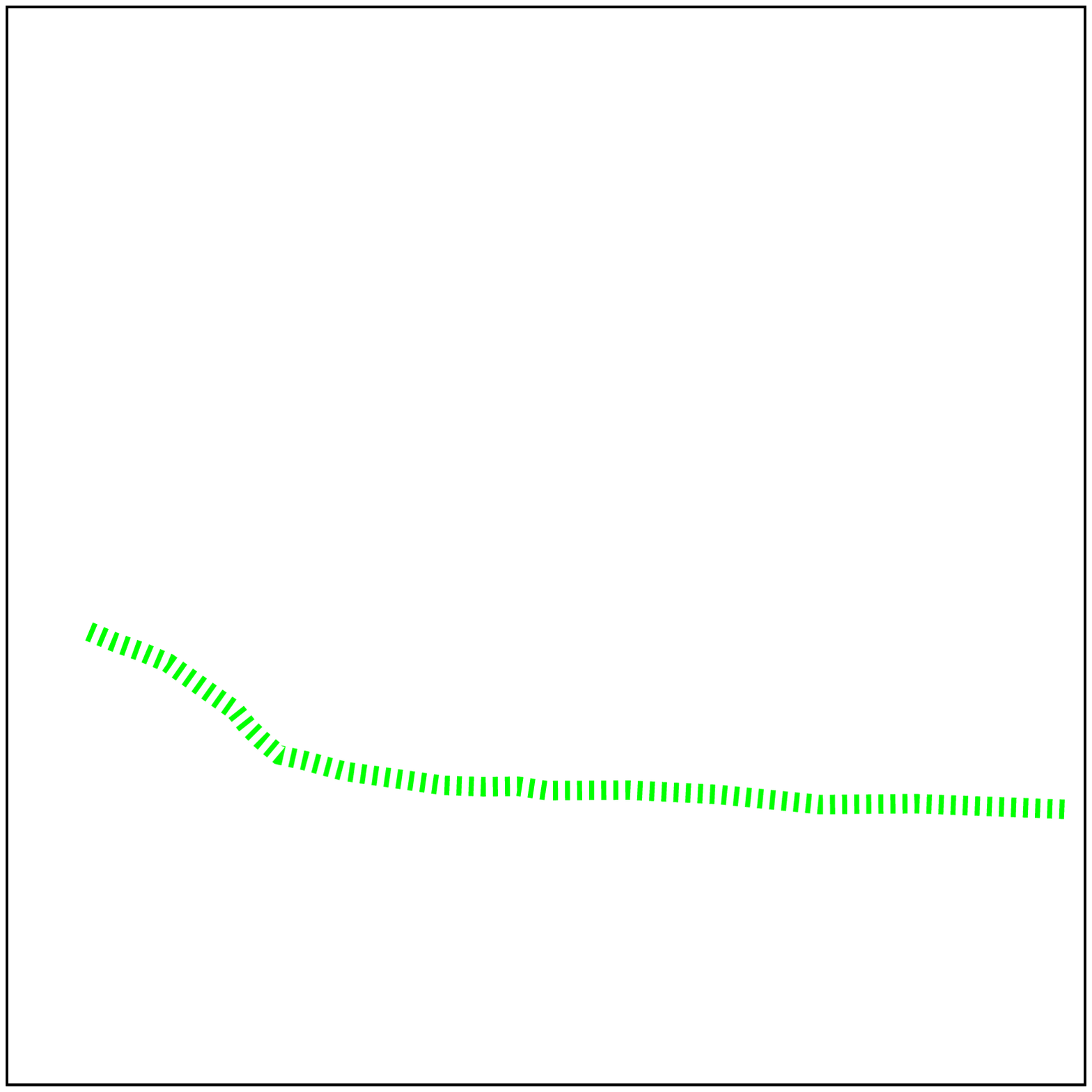}}
  \put(2.4,8.2){\includegraphics[angle=270,width=0.7\textwidth]{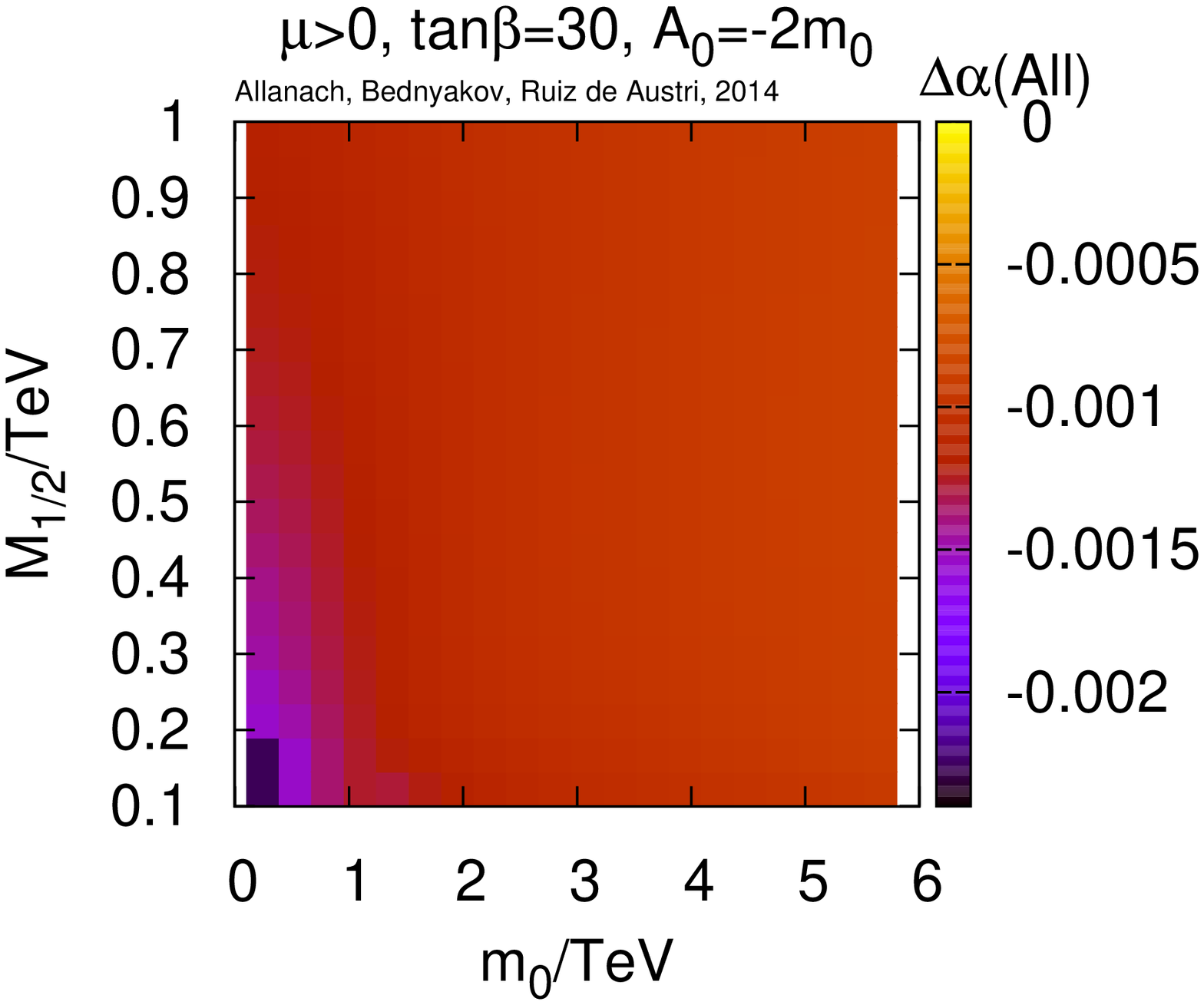}}
  \put(3.22,7.64){\includegraphics[angle=270,width=0.45\textwidth]{atlasExcl}}
  \put(0,7.8){(a)}
  \put(3.3,7.8){(b)}
  \put(0,5){(c)}
  \put(3.3,5){(d)}
  \put(0,2.2){(e)}
  \put(3.3,2.2){(f)}
\end{picture}
\end{center}
\caption{\label{fig:cmssmUn} Relative effect of highest order terms on
  unification in the CMSSM. The CMSSM 
  parameters in (a) and (b) coincide with the latest ATLAS searches for jets
  and missing 
  energy interpreted in the 
  CMSSM~\cite{Aad:2014wea}. On the colour legend we have
  labeled the default {\tt SOFTSUSY}\ calculation by (None) and the one
  including the higher order corrections by (All).
  The regions in (a) and (b) below the dashed line are excluded at the 95$\%$
  confidence level by at least one of the most restrictive ATLAS jets plus
  missing energy searches.}
\end{figure}
We shall now display some of our results in the parameter plane recently
defined by ATLAS.  
We have combined the two most restrictive exclusion limits from their direct
LHC 7/8 
TeV searches~\cite{Aad:2014wea} into one exclusion limit: if a
point is excluded at 95$\%$ confidence level by either or both of them, we
count it as excluded.  Comparing Figs.~\ref{fig:cmssmUn}a,\ref{fig:cmssmUn}b,
we see that the higher order corrections introduce a constant term which 
makes $\alpha_3(M_{GUT})$ approximately 0.001 below
$\alpha_1(M_{GUT})=\alpha_2(M_{GUT})$. 
When studying Yukawa unification, we diverge from the ATLAS plane and instead
vary two parameters that control Yukawa unification more directly: $\tan
\beta$ and $A_0$. We fix $m_0=2$ TeV, $M_{1/2}=0.6$ TeV so that, at $\tan
\beta=30$, the point is allowed (as shown by reference to
Fig.~\ref{fig:cmssmUn}a,b). It is expected~\cite{Allanach:2011ut} that the LHC
limits should 
only be weakly 
dependent upon $\tan \beta$ and so we expect this $\tan \beta$, $A_0$
plane to not be excluded by
them. 
Figs~\ref{fig:cmssmUn}c,\ref{fig:cmssmUn}d show that
the difference in bottom and tau Yukawa couplings doesn't change much
in the region closest to Yukawa unification at around $\tan \beta \approx 2$:
less than half a percent (in the less unified direction) is accounted for by
the higher order corrections there.
However, bottom-top Yukawa unification is made slightly better, by a few 
percent or so (seen from Figs~\ref{fig:cmssmUn}e,\ref{fig:cmssmUn}f), but 
even at high values of 
$\tan \beta$, there is a 10$\%$ discrepancy between the two couplings. One
would need quite large GUT scale threshold corrections to explain this
sizable discrepancy.

\begin{figure}
\unitlength=1in
\begin{center}
\begin{picture}(6,7.8)
  \put(-0.75,2.8){\includegraphics[angle=270,width=0.7\textwidth]{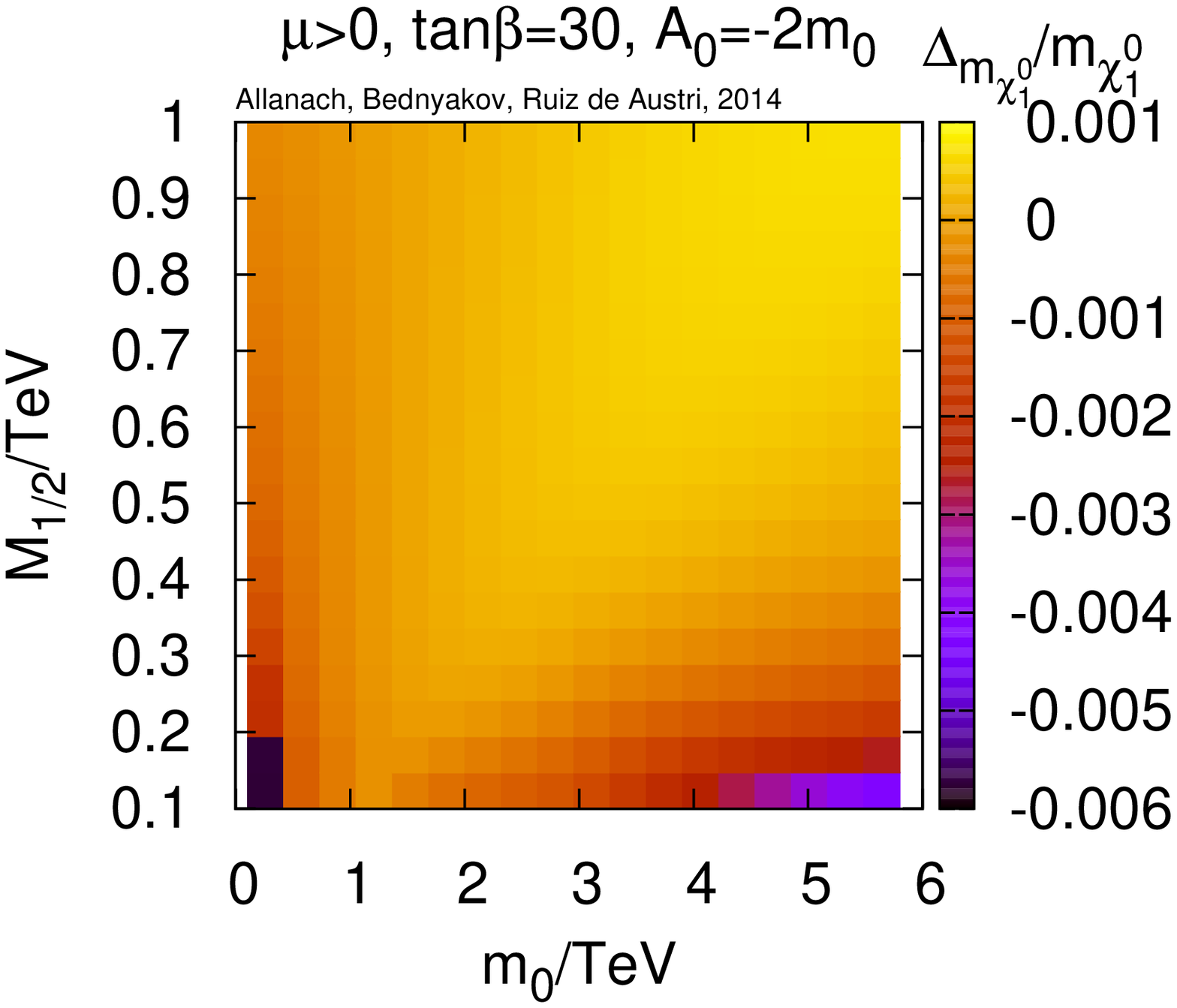}}
  \put(0.08,2.24){\includegraphics[angle=270,width=0.45\textwidth]{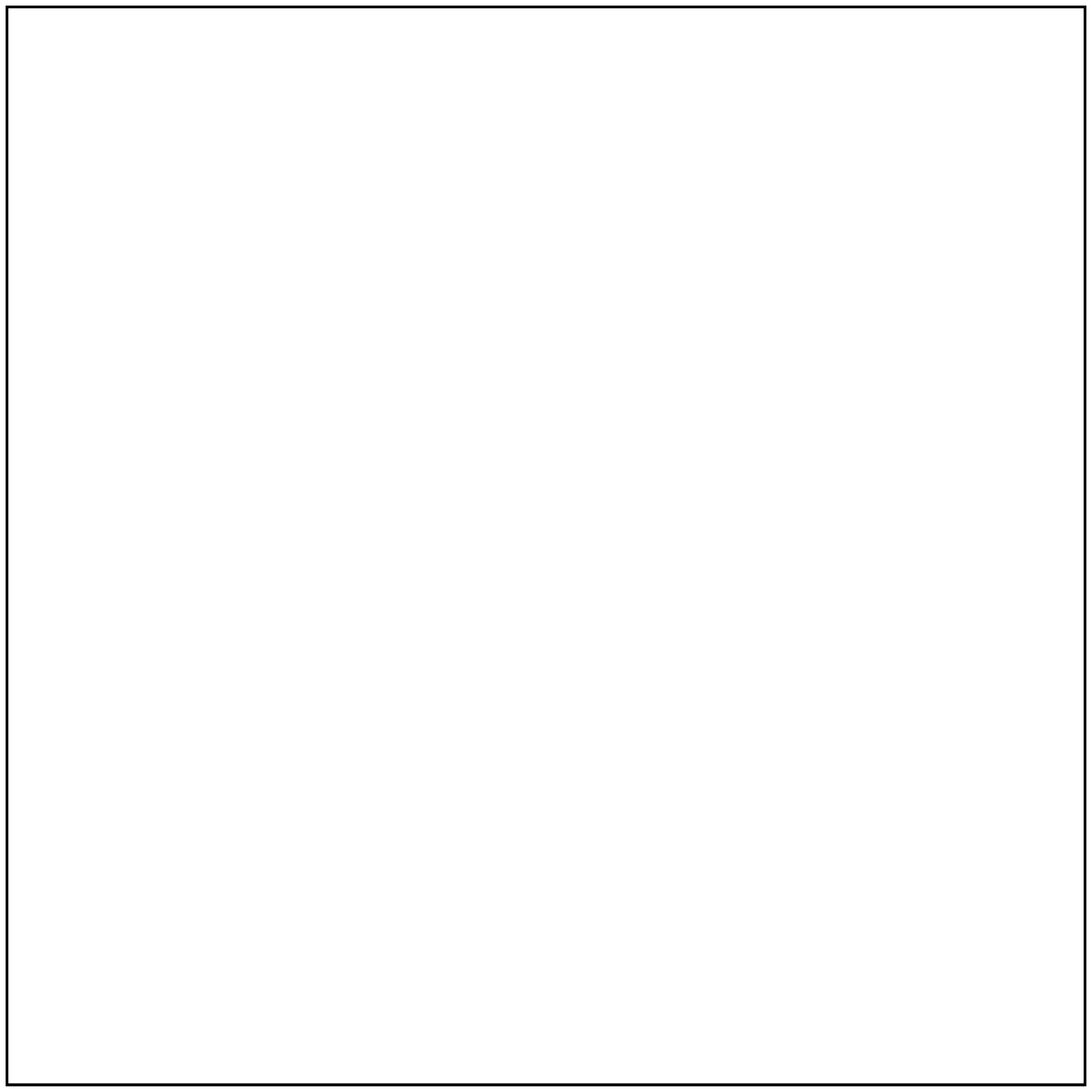}}
  \put(0.08,2.24){\includegraphics[angle=270,width=0.45\textwidth]{atlasExcl}}
  \put(-0.75,5.5){\includegraphics[angle=270,width=0.7\textwidth]{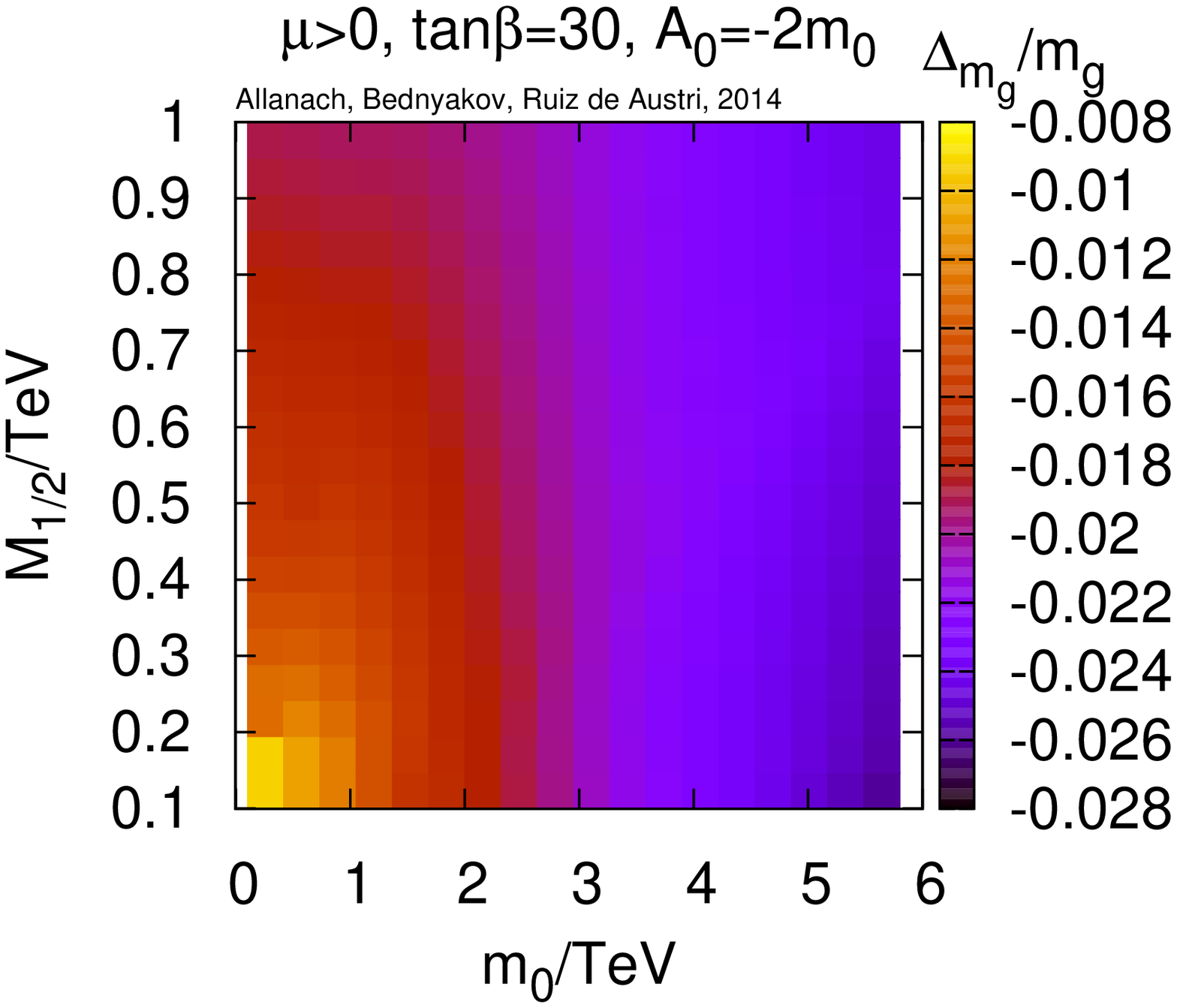}}
  \put(0.08,4.94){\includegraphics[angle=270,width=0.45\textwidth]{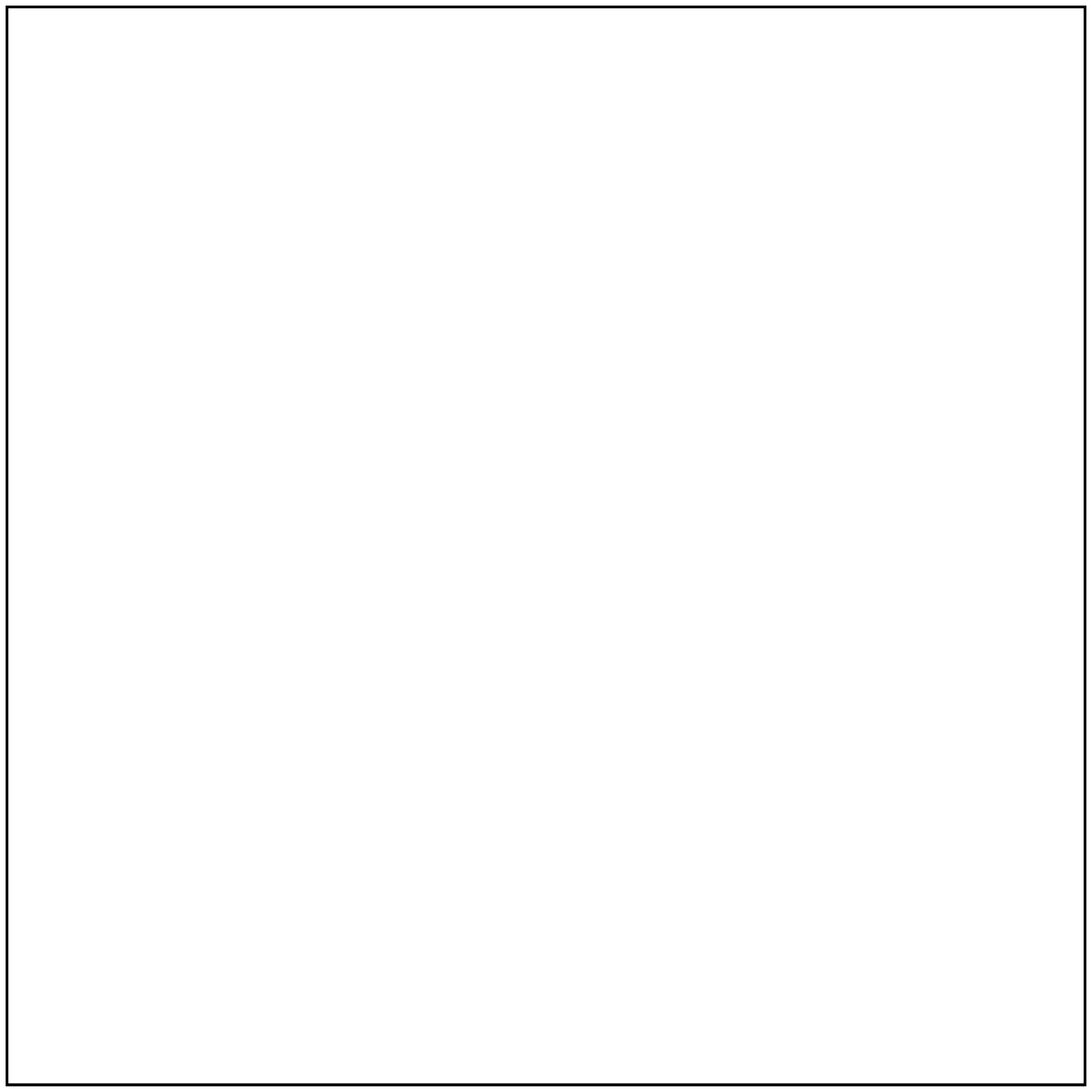}}
  \put(0.08,4.94){\includegraphics[angle=270,width=0.45\textwidth]{atlasExcl}}
  \put(2.4,2.8){\includegraphics[angle=270,width=0.7\textwidth]{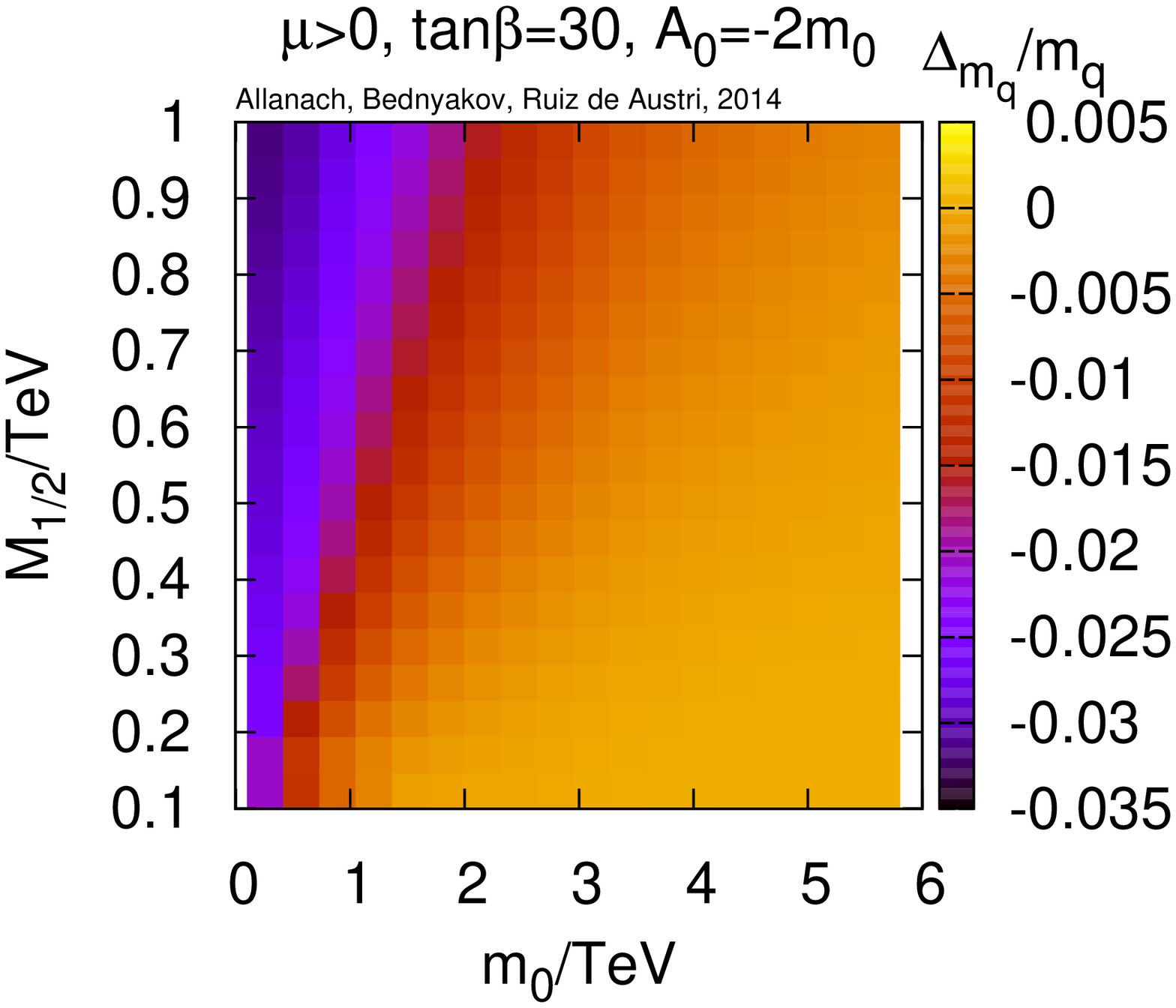}}
  \put(3.22,2.24){\includegraphics[angle=270,width=0.45\textwidth]{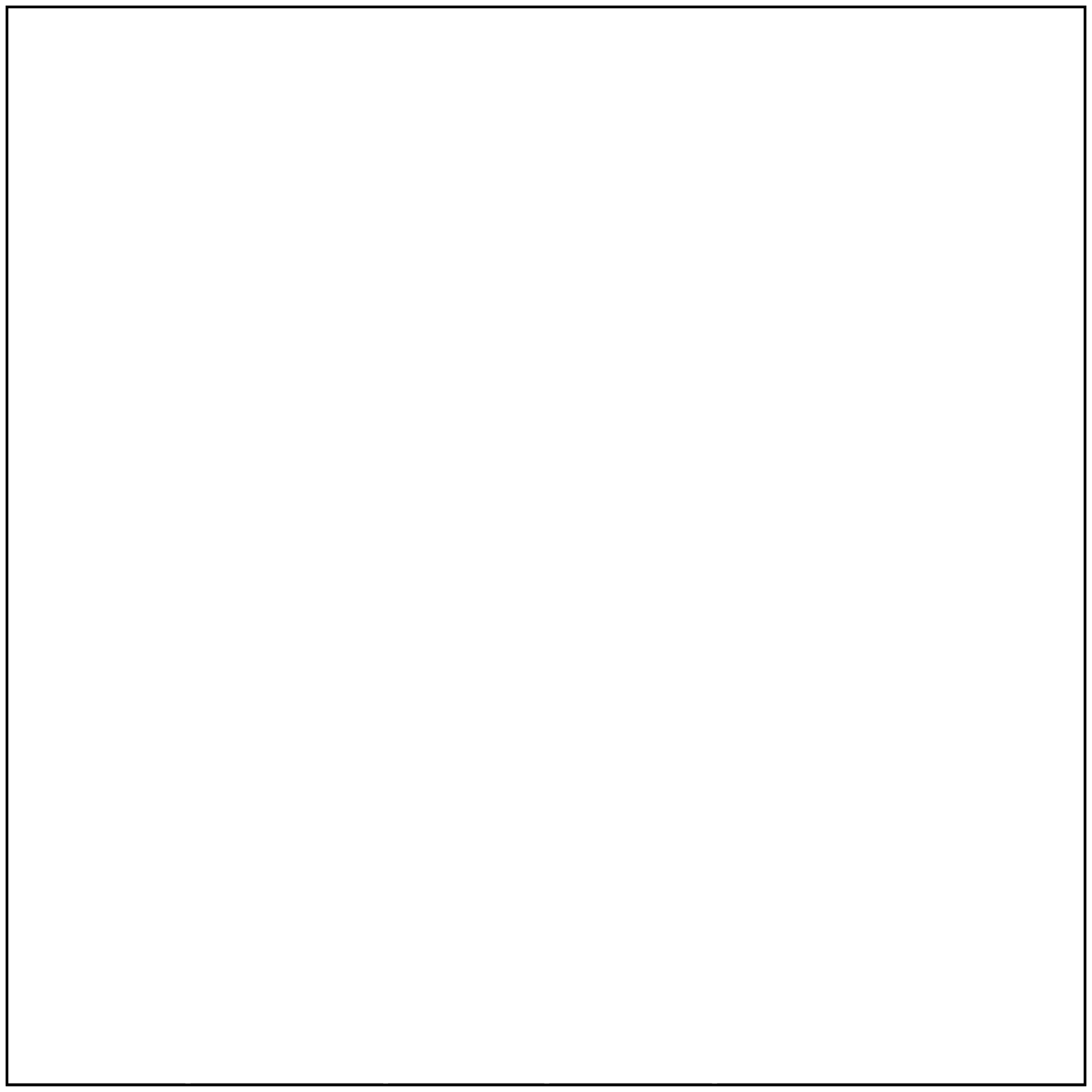}}
  \put(3.22,2.24){\includegraphics[angle=270,width=0.45\textwidth]{atlasExcl}} 
\put(2.4,5.5){\includegraphics[angle=270,width=0.7\textwidth]{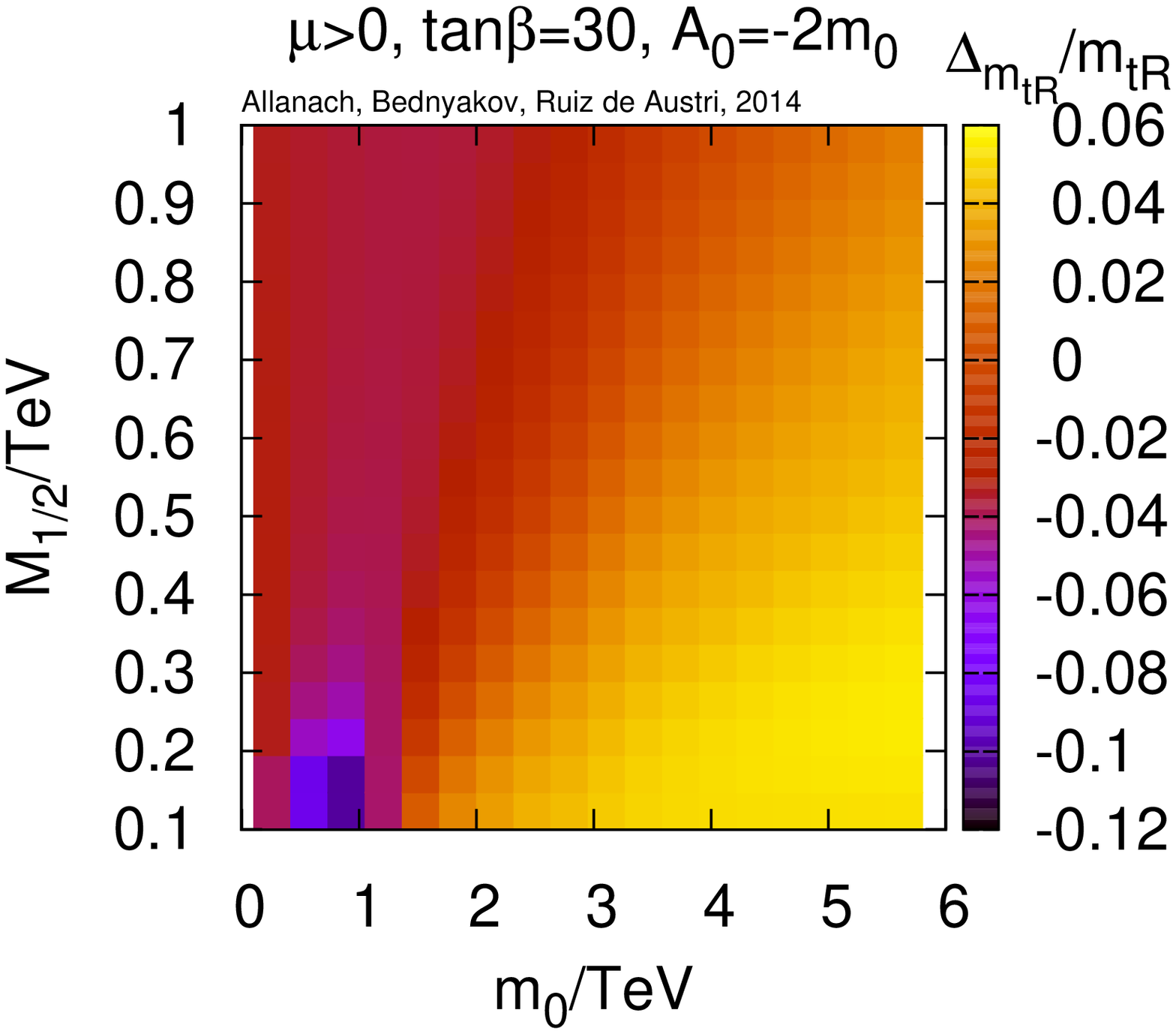}}
  \put(3.22,4.94){\includegraphics[angle=270,width=0.45\textwidth]{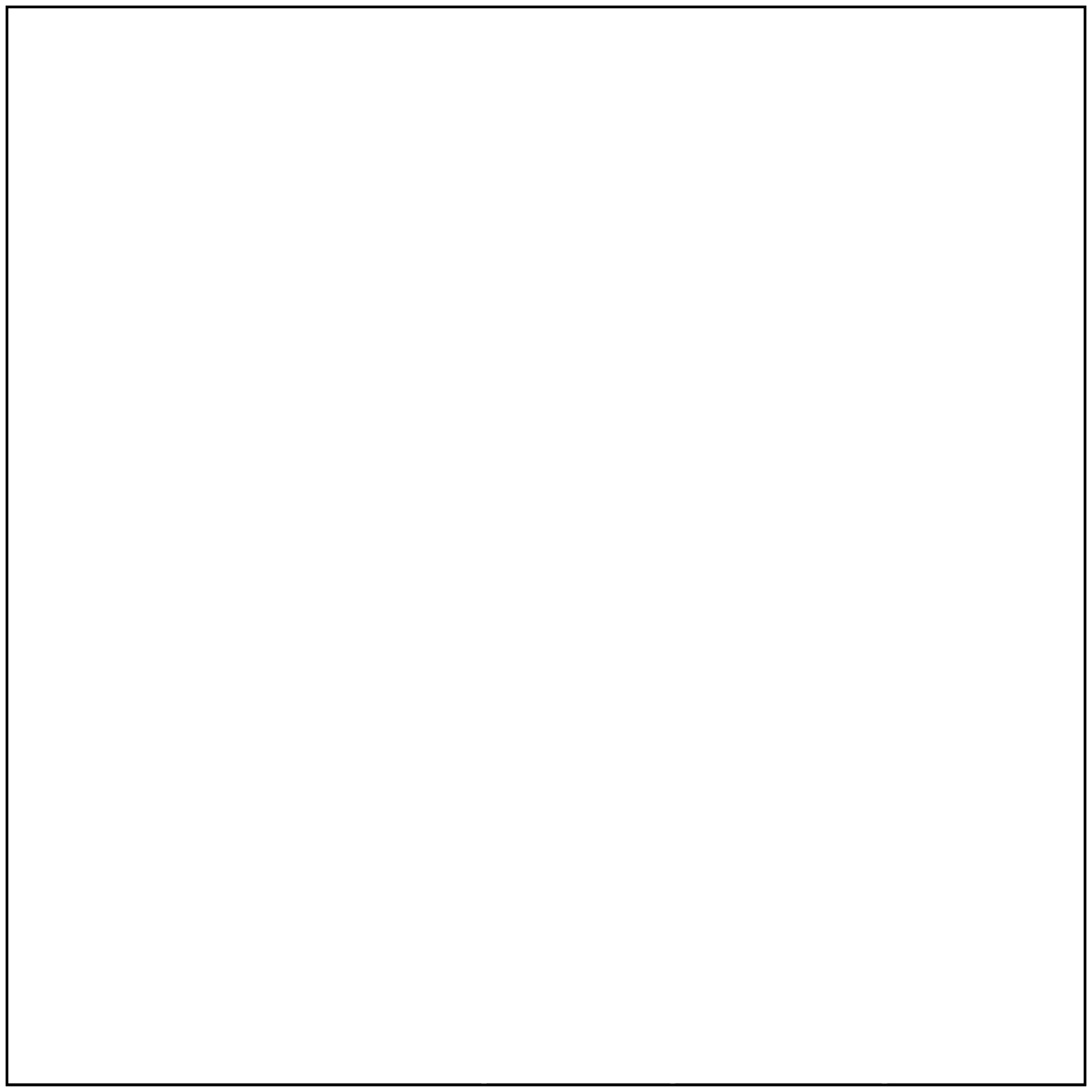}} 
  \put(3.22,4.94){\includegraphics[angle=270,width=0.45\textwidth]{atlasExcl}}
  \put(-0.75,8.2){\includegraphics[angle=270,width=0.7\textwidth]{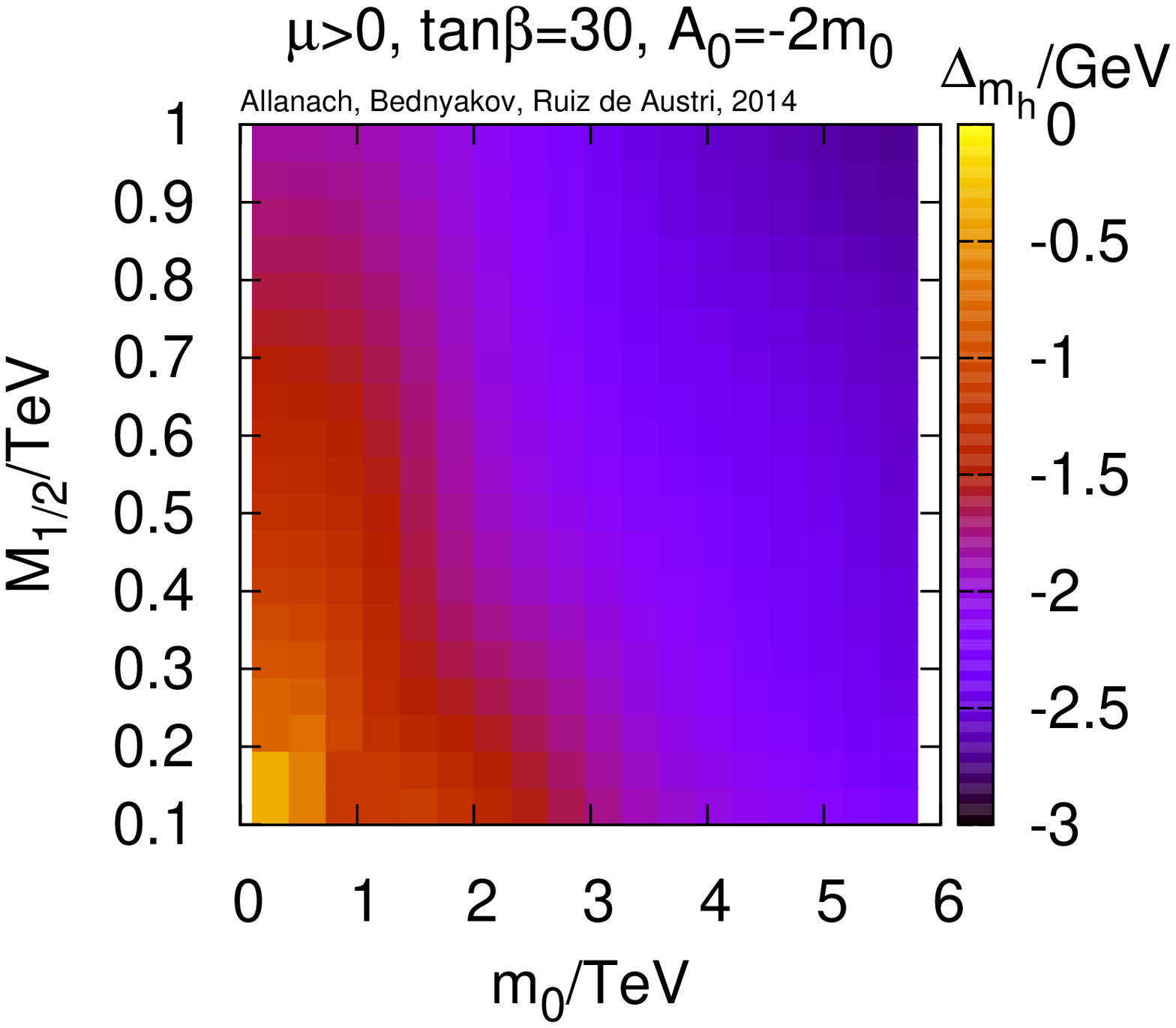}}
  \put(0.08,7.64){\includegraphics[angle=270,width=0.45\textwidth]{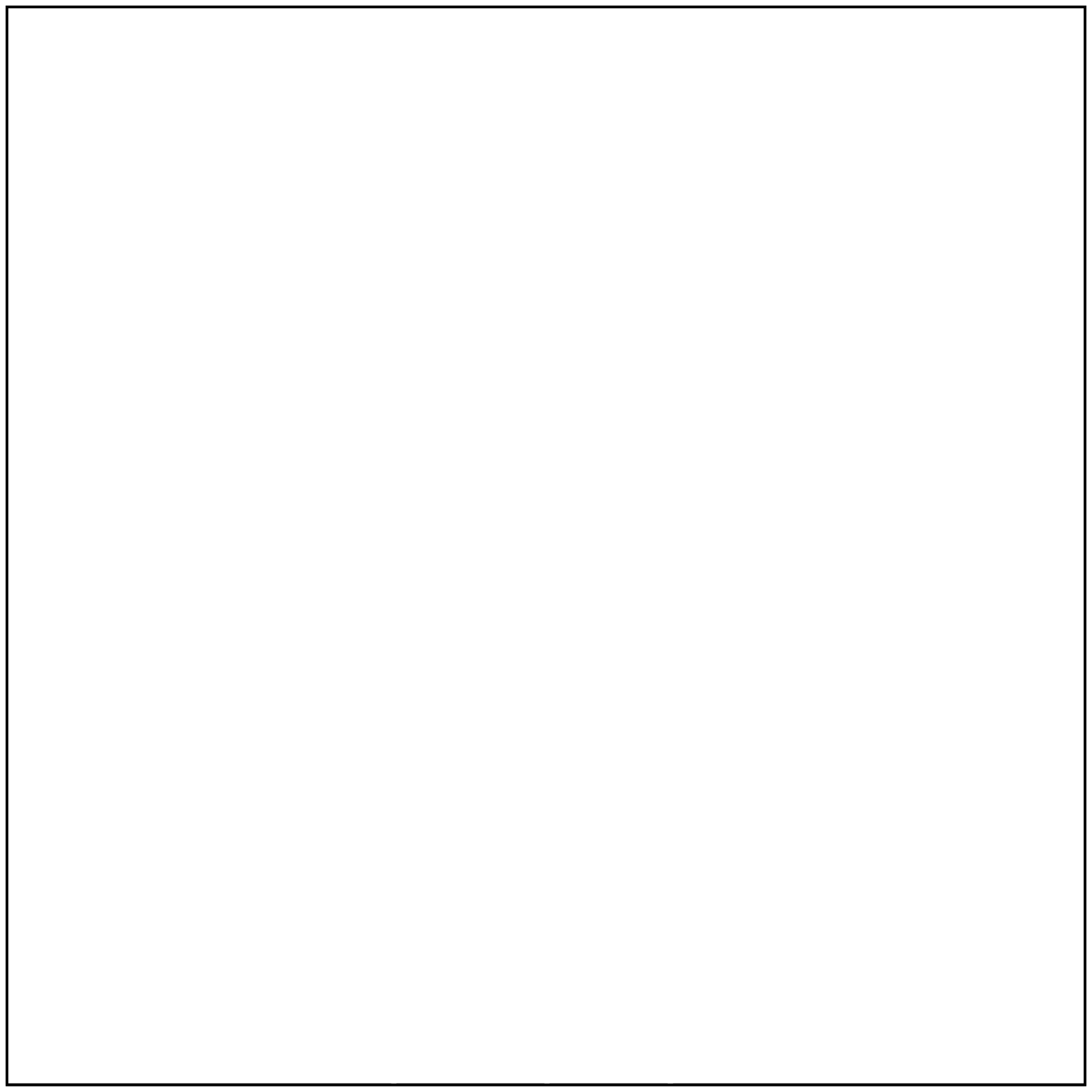}}
  \put(0.08,7.64){\includegraphics[angle=270,width=0.45\textwidth]{atlasExcl}}
  \put(2.4,8.2){\includegraphics[angle=270,width=0.7\textwidth]{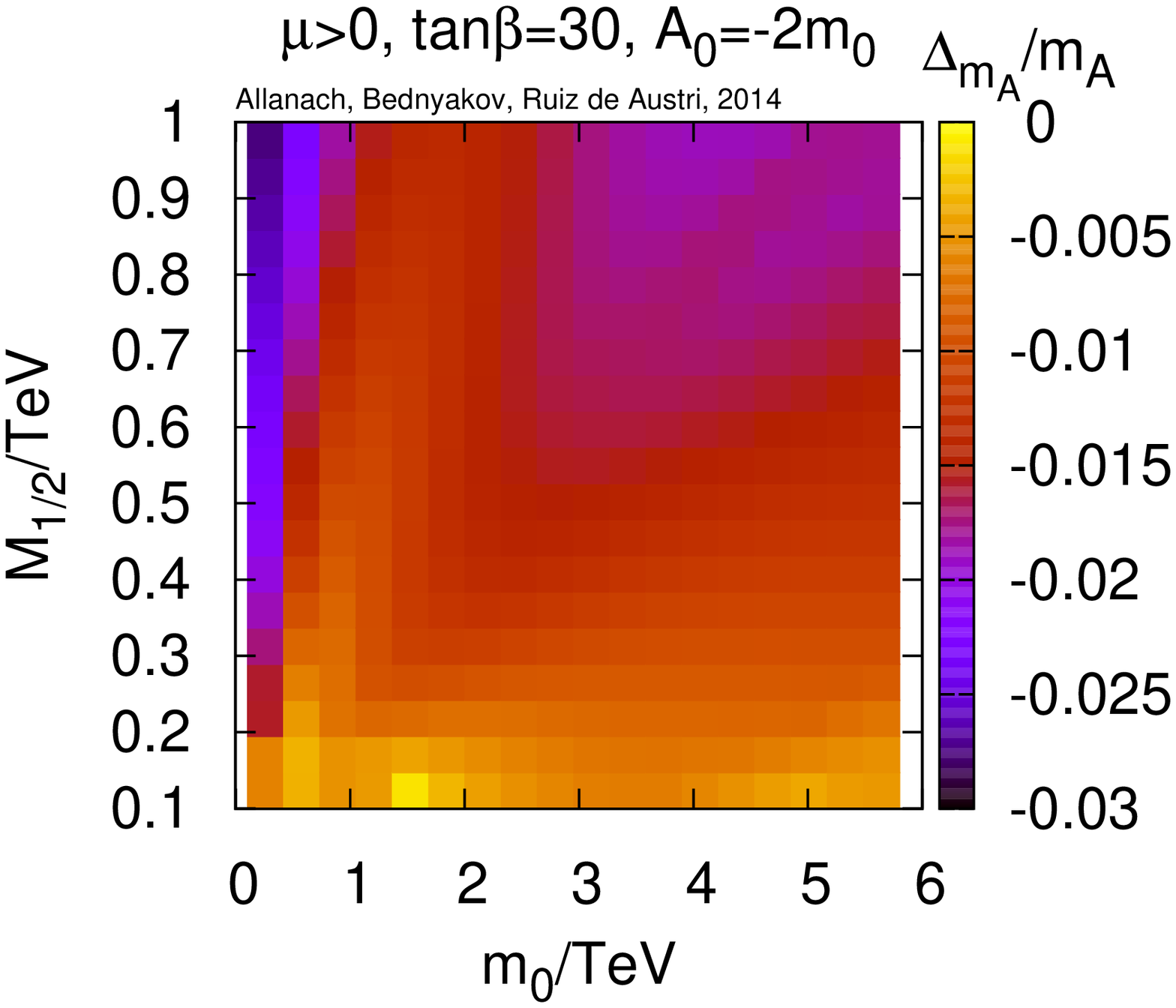}}
  \put(3.22,7.64){\includegraphics[angle=270,width=0.45\textwidth]{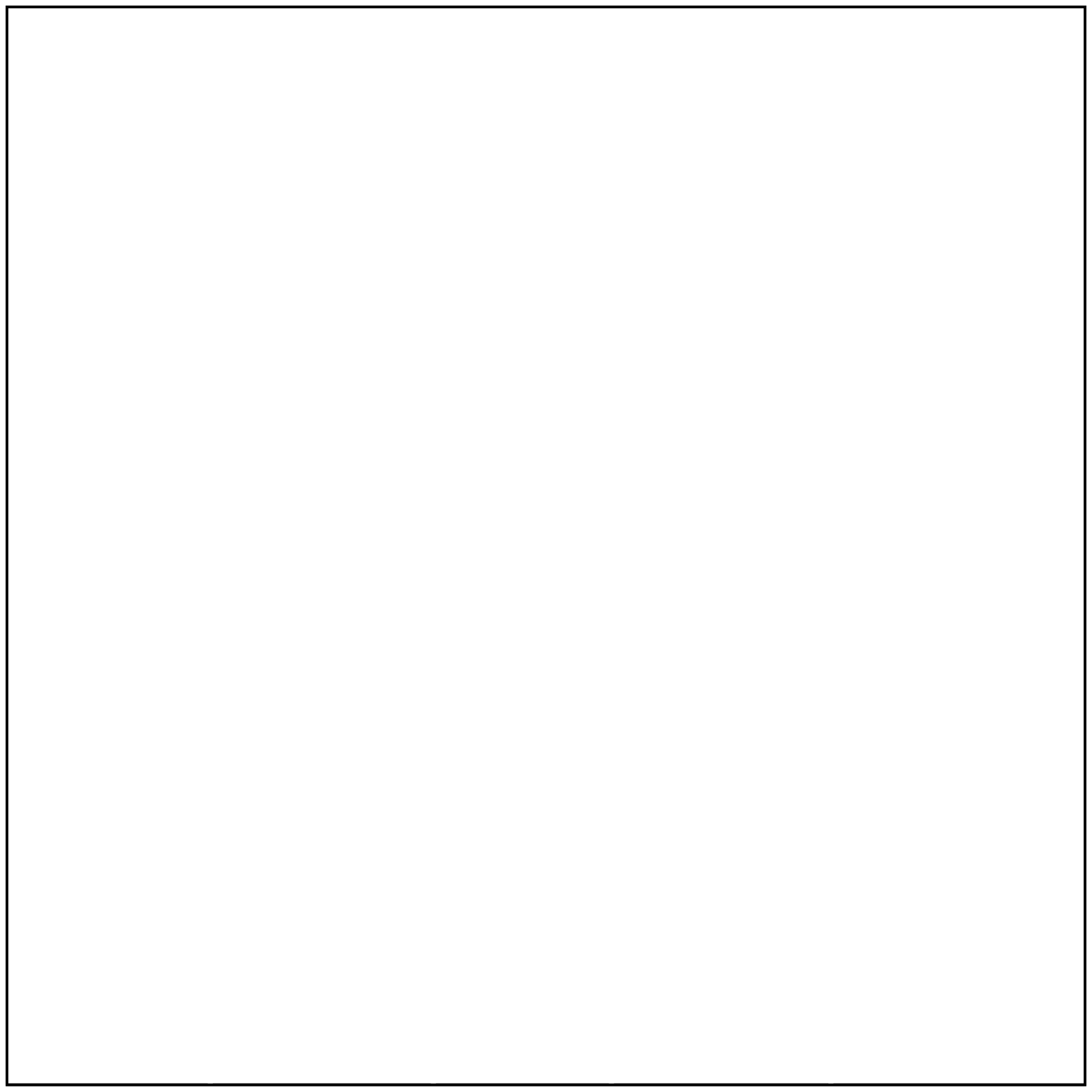}}
  \put(3.22,7.64){\includegraphics[angle=270,width=0.45\textwidth]{atlasExcl}}
  \put(0,7.8){(a)}
  \put(3.2,7.8){(b)}
  \put(0,5){(c)}
  \put(3.2,5){(d)}
  \put(0,2.2){(e)}
  \put(3.2,2.2){(f)}
\end{picture}
\end{center}
\caption{\label{fig:cmssm} Relative effect of highest order terms (three-loop
  RGEs for gauge and Yukawa couplings and two-loop threshold corrections to
  third family fermion masses and $g_3$) on various
  particle pole masses in the CMSSM. The CMSSM 
  parameters coincide with a parameter plane where limits from the latest
  ATLAS searches for jets and missing 
  energy were presented in the 
  CMSSM~\cite{Aad:2014wea}.
  Solid contours of iso-mass calculated including all of our higher order
  corrections 
  are   overlayed on each 
  figure, with each contour labelling the mass in GeV. $\Delta m/m$ denotes
  the change that was induced by the higher order corrections and is shown as
  the background colour in each plot. 
  Here, $m_g$ denotes the
  gluino mass and $m_q$ the average squark mass from the first two 
  generations. The region below the dashed line is excluded at the 95$\%$
  confidence level by at least one of the most restrictive ATLAS jets plus
  missing energy searches.}
\end{figure}
In Fig.~\ref{fig:cmssm}, we show some contours of important MSSM particle
masses as 
well as their relative change due to the higher order corrections. The region
below the dashed line is excluded by either one or both of the most
restrictive ATLAS SUSY searches~\cite{Aad:2014wea}. 
We see that the gluino, the lightest CP even Higgs mass, the first two
generation average squark mass and
the CP odd Higgs mass $m_A$ typically become reduced by
1-3$\%$ by the higher order corrections in the region allowed by the search. 
In the CMSSM, the dominant production of SUSY particles is via gluino and 
squark production. 
The mass of the lightest neutralino changes less: typically at the one
per-mille level, whereas the lightest stop mass has larger contributions from
the higher order corrections: up to about $\pm 8\%$. CMSSM signatures involving
the lightest stops are therefore more sensitive to the higher order
contributions. 
\begin{figure}
\unitlength=1in
\begin{center}
\begin{picture}(3,3)
  \put(-0.75,2.8){\includegraphics[angle=270,width=0.7\textwidth]{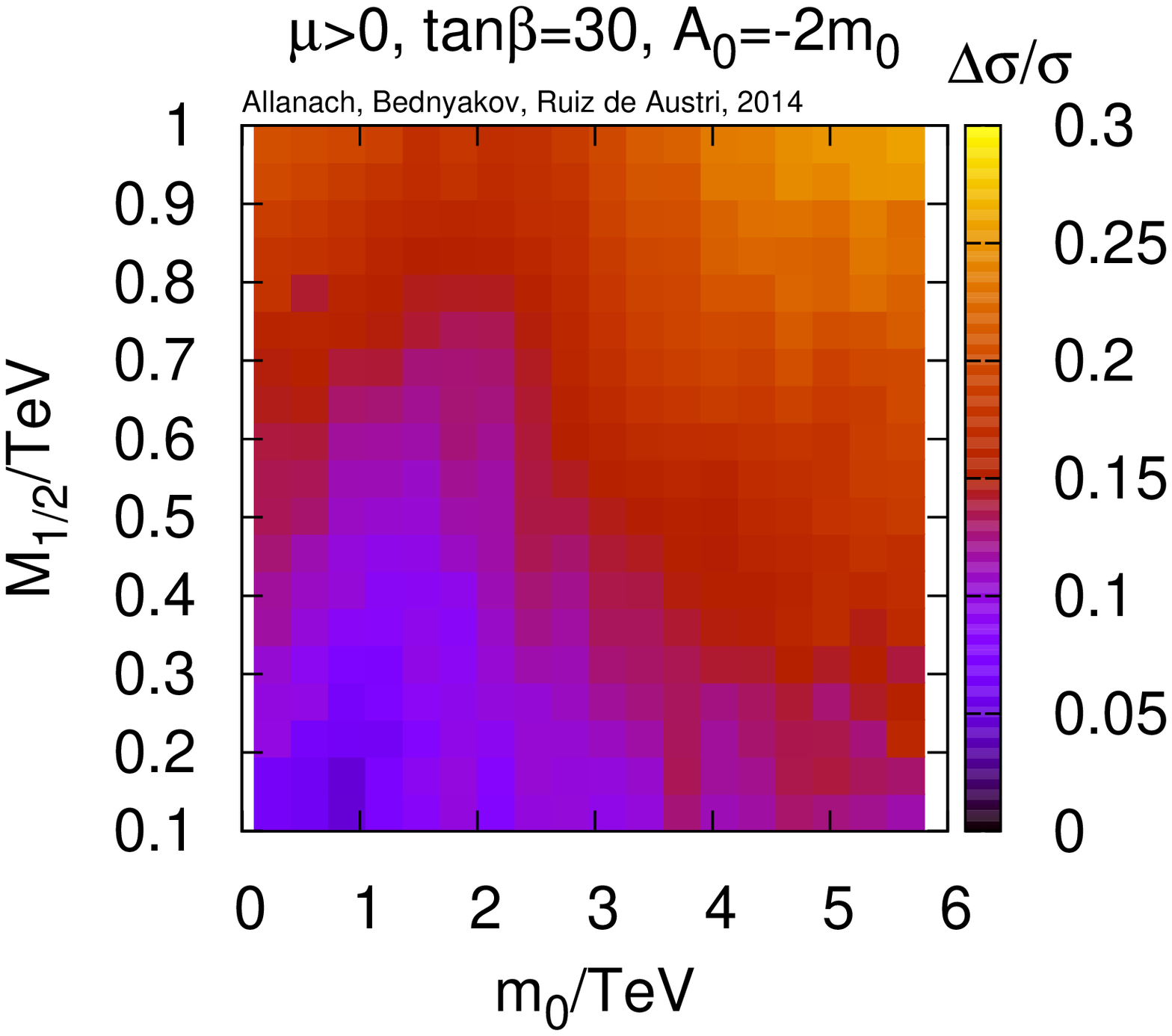}}
  \put(0.08,2.24){\includegraphics[angle=270,width=0.45\textwidth]{atlasExcl}}
\end{picture}
\end{center}
\caption{\label{fig:sig} Relative effect of highest order terms (three-loop
  RGEs for gauge and Yukawa couplings and two-loop threshold corrections to
  third family fermion masses and $g_3$) on the predicted LHC 
  SUSY production cross-section. The CMSSM 
  parameters coincide with the latest ATLAS searches for jets and missing
  energy interpreted in the 
  CMSSM~\cite{Aad:2014wea}. The region below the dashed line is excluded at
  the 95$\%$ 
  confidence level by at least one of the most restrictive ATLAS jets plus
  missing energy searches.} 
\end{figure}
The reduction of gluino and squark masses makes the SUSY
production cross-section larger. 
As Fig.~\ref{fig:sig} shows, this results in an increase of 10-26$\%$ in the
cross-section within
the region not excluded by current searches. This is therefore our estimate
for the theoretical uncertainty upon the next-to-leading order 
cross-section induced from 
spectrum uncertainties (note however that it does not include theoretical
uncertainties coming from the next-to-next-to leading order cross-section,
however that can easily be 
obtained from scale dependence in the next-to-leading order result).

\begin{figure}
\unitlength=1in
\begin{center}
\begin{picture}(3,3)
  \put(-0.75,2.8){\includegraphics[angle=270,width=0.7\textwidth]{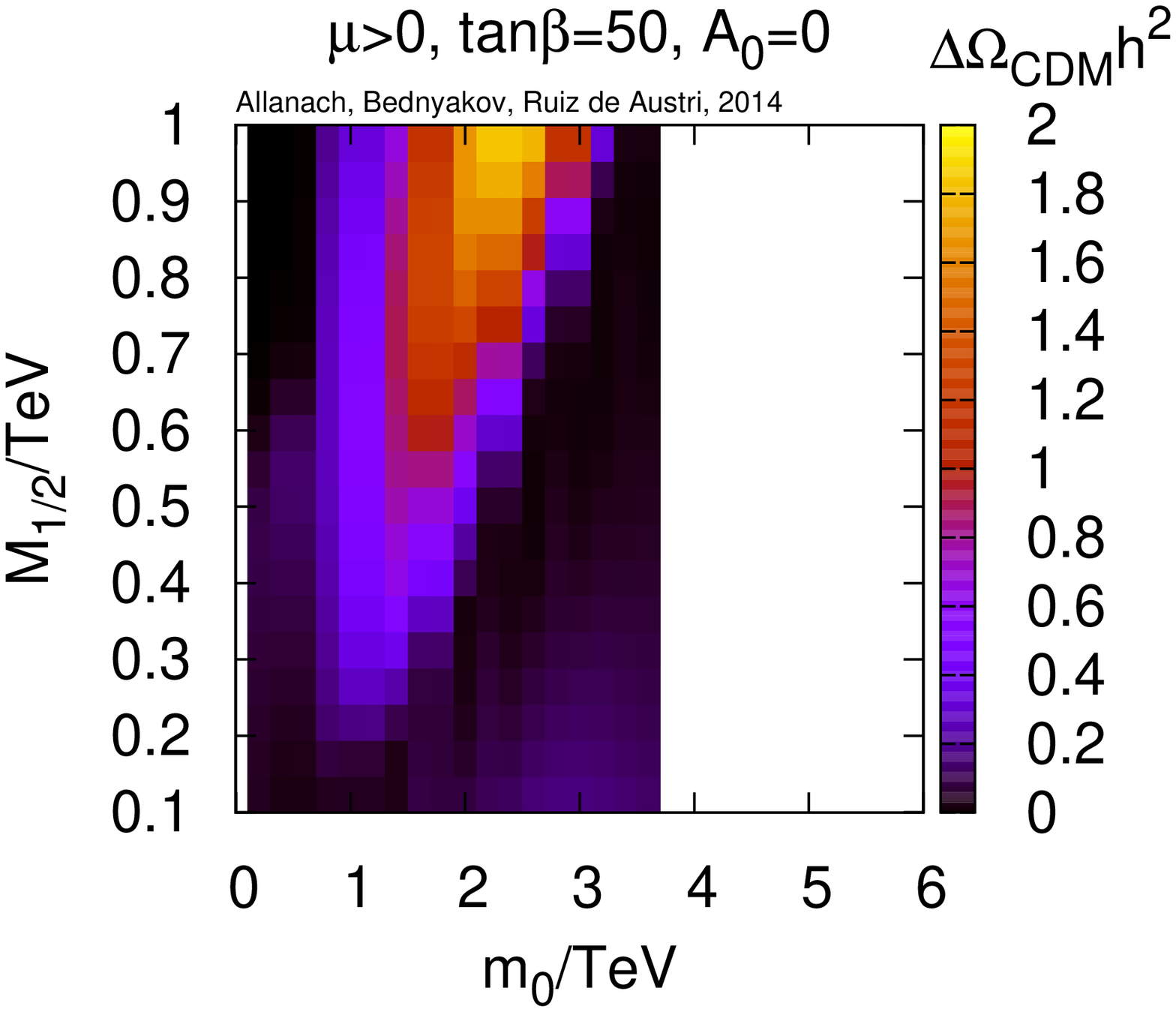}}
  \put(0.08,2.24){\includegraphics[angle=270,width=0.45\textwidth]{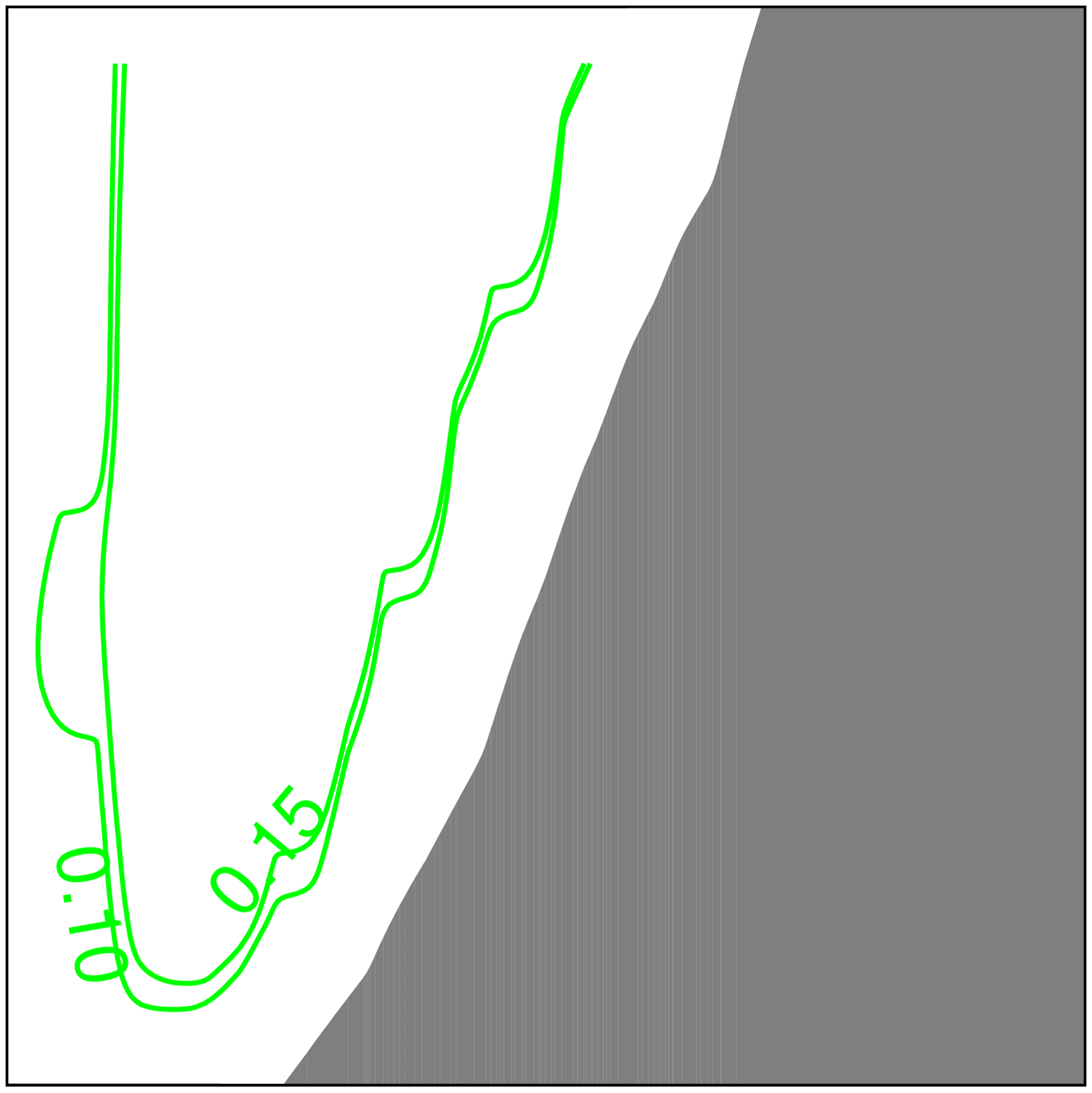}}
\end{picture}
\end{center}
\caption{\label{fig:dm} Effect of highest order terms (three-loop
  RGEs for gauge and Yukawa couplings and two-loop threshold corrections to
  third family fermion masses and $g_3$) on the predicted dark matter relic
  density in the CMSSM in a high $\tan \beta$ scenario. Contours of iso-relic
  density for the highest order prediction are overlayed. Also shown is the
  change in the position of the border of successful electroweak symmetry
  breaking: the white  region 
  (and to the right) is ruled out for higher loop corrections, whereas
  the lighter one is ruled out according to the standard {\tt SOFTSUSY3.4.1}
  calculation.} 
\end{figure}
We next perform a scan at high $\tan \beta$, displaying a region where the
dark matter relic abundance appears to have the correct value compared to the
value inferred from
cosmological observationals. 
In Fig.~\ref{fig:dm}, we show this as the region between the two green
contours. On the other hand, the background colour shows the apparent induced 
theoretical uncertainty in the prediction of the dark matter relic density
from our higher order terms. We have defined
$\Delta \Omega_{CDM} h^2$ to be the `$\Delta$ All'$-$`None' value. 
We see that, for $m_0>1$ TeV, it is swamped by theoretical uncertainties and
the prediction is completely unreliable. This is not unexpected at the focus
point, because of huge sensitivities to $Y_t$~\cite{Allanach:2012qd}. While
the uncertainties for 
fixed CMSSM parameters are huge, it is true that the region of dark
matter relic abundance that agrees with observations will be present
somewhere. However, it may move significantly with $m_0$. The contours shown
track to be close to the boundary of successful electroweak symmetry breaking,
shown by the white region. As we move across the plot from
left to right, the value of $\mu^2(M_{SUSY})$ as predicted by minimisation of
the Higgs potential decreases, and finally becomes less than zero in the white
region, signalling incorrect electroweak symmetry breaking. 
If we omit the
higher order corrections, we obtain instead the grey region, which would have
its own contours of dark matter relic density predicted to be compatible with
observations. 

\section{Summary and Conclusions}
 We have incorporated full three-loop MSSM $R-$parity and CP-conserving
 conserving RGEs as well as some leading two-loop 
 threshold corrections to the QCD gauge coupling and third family fermion
 masses into the {\tt SOFTSUSY} spectrum calculator. The corrections included
 are: $\mathcal O(\alpha_s^2)$ corrections to $m_t$, $\mathcal O(\alpha_s^2)$,
 and $\mathcal O(\alpha_s \alpha_{t,b})$ 
 corrections to $\alpha_s$, and $\mathcal O(\alpha_s^2)$, 
 $\mathcal O(\alpha_s \alpha_{t,b})$, $\mathcal O(\alpha_{t,b}^2)$, $\mathcal O(\alpha_t \alpha_b)$, and $\mathcal O(\alpha_{t,b} \alpha_\tau)$ corrections to
 $m_b$. $\mathcal O(\alpha_\tau^2), \mathcal O(\alpha_{t,b} \alpha_\tau)$ 
 corrections to $m_\tau$ are also included. These corrections make gauge and
 Yukawa unification predictions more accurate. 
 We report up to a 3 GeV  (usually {\em negative}) change in
 the prediction  of  the lightest CP even Higgs mass, mostly due to
 the reduction of the running top Yukawa coupling originating from the two-loop threshold correction
 to the top quark mass.
 This result looks complementary to that reported in Refs.~\cite{Kant:2010tf,Feng:2013tvd} --
 three-loop diagrammatic corrections $\mathcal{O}(\alpha_t \alpha_s^2)$ 
 to the lightest Higgs mass are estimated to be of the same order but  {\em positive}.
 One could envisage some cancellations between these kind of corrections if combined 
 in a single calculation.
 
 The inclusion of the higher order terms also gives a good estimate
 for the size of theoretical uncertainties in the sparticle mass predictions
 from higher order corrections. Some sparticle masses have 10$\%$
 uncertainties when running to and from the GUT scale, as in the CMSSM, where
 small threshold effects become amplified by sensitive renormalisation group
 running. On the other hand, in the pMSSM, where there is only running between
 $M_{SUSY}$ and $M_Z$, the theoretical uncertainties in sparticle masses are
 smaller: typically at the one percent level. The uncertainties in the spectrum
 have a knock-on effect on derived observables: for example, the predicted
 relic density of dark matter, since it depends so sensitively on sparticle
 masses in some parts of parameter space, can have order 1 relative theoretical
 uncertainties. The total LHC sparticle production cross-section can have a
 $30\%$ error in the CMSSM (this decreases to a percent or so in the pMSSM).
 The change in the running
 value of the top quark mass induces a particularly large change in the
 higgsino mass parameter $\mu$ at the focus point of the CMSSM at large $m_0$,
 resulting in 
 huge theoretical uncertainties in some neutralino and chargino masses. 
 This is thus an important input for global fits of the CMSSM (see, for example
Refs.~\cite{Buchmueller:2011ab,Allanach:2011wi,Roszkowski:2014wqa,Fowlie:2013oua}). It 
is probable that regions of parameter space at high 
$m_0$ are weighted 
incorrectly. Ideally, the fit would be performed including a calculated
 theoretical error (particularly that coming from the dark matter relic
 density constraint~\cite{Belanger:2005jk}). This could come from estimating
 the corrections using  
 our higher order corrections in order to quantify the uncertainty, or from
 renormalisation scale dependence of observables (for instance, how much would
 the dark matter relic density prediction change if $M_{SUSY}$ were varied by
 a factor of 2?)

 Neither $\mathcal O(\alpha_s \alpha_t) $ nor $\mathcal O(\alpha_t^2)$
 corrections are included in the calculation of the running value of
 $m_t$. Since parts of the phenomenology are so 
 sensitively dependent upon the precise value of of $m_t$ in parts of
 parameter space (especially the focus point of the CMSSM), an important
 future work will be to include these. We estimate that current uncertainties
 on the extreme focus point region are huge, and need to be decreased by the
 calculation and addition of these terms. 
 We note that currently, no other publicly supported spectrum calculator
 contains our higher 
 order terms. There has been a tendency in the recent literature for some
 authors to increase the SUSY breaking mass scales to several tens of TeV, or
 even higher. 
 In this case, to get a $m_h$ prediction that is very accurate, the fixed
 order calculations employed in {\tt SOFTSUSY} could be subject to corrections
 of several GeV~\cite{Draper:2013oza,Bagnaschi:2014rsa}. For a more
 accurate prediction, additional log re-summation 
 should be implemented: another important possible future direction for
 research. 

 There has been attention in the literature 
 on the question of whether full top-bottom-tau Yukawa
 unification is possible in supersymmetric minimal SO(10) GUT
 models while respecting current
 data~\cite{Blazek:2002ta,Altmannshofer:2008vr,Anandakrishnan:2014nea}.  
 It will be an interesting future project to examine to what extent this is
 possible or not while including the important effects of the higher order
 corrections, although this should only be done after the inclusion of
 $\mathcal O(\alpha_s \alpha_t) $ and $\mathcal O(\alpha_t^2)$ corrections to
 $m_t$. 
 A more formidable future enterprise would
 be to  include direct two-loop threshold corrections to sparticle masses (e.g., 
 with the help of TSIL \cite{Martin:2005qm} package~\cite{Martin:2005ch,Martin:2006ub}). 
 The corrections that we have included are necessary if such corrections are to
 increase the accuracy of sparticle mass predictions.  
 In addition to this,  leading three-loop corrections \cite{Martin:2007pg,Kant:2010tf} to the
 lightest CP-even Higgs boson could be incorporated to further reduce the
 uncertainty of  the corresponding mass prediction.

 We have examined the effects of the higher order terms that we include
 upon apparent discrepancies in various predictions of unification at the GUT
 scale. We fix gauge and Yukawa couplings to data at $M_Z$, assuming some
 value of $\tan \beta$. Then, by evolving to $M_{GUT}$, where the electroweak
 gauge couplings meet, we obtain GUT scale gauge and Yukawa couplings. The
 discrepancy between $\alpha_3(M_{GUT})-\alpha_1(M_{GUT})$ is typically larger 
 once higher order corrections are included (particularly the two-loop
 threshold corrections to $\alpha_3(M_Z)$). However, it is in any case only at
 the per mille level and can easily be explained by small GUT threshold
 corrections. Yukawa unification has larger apparent GUT-scale discrepancies in
 generic parts of parameter space. It is affected mostly by higher order top
 mass and $\alpha_3(M_Z)$ threshold corrections. We have studied examples
 where these change the GUT-scale
 discrepancies by 4$\%$. This would certainly have an impact on detailed GUT
 model building, in order to explain the discrepancy with, for example,
 GUT-scale  threshold corrections. 
 
 We have provided details in the Appendix of how to compile and run a new
 publicly available version of {\tt    SOFTSUSY} that incorporates the higher
 order terms discussed above\footnote{There may be minimal changes (from a
   user's 
 perspective) to this procedure in future versions where additional
 corrections are added.}. We hope that this provision will aid other studies of
 unification and quantification of theoretical uncertainties in the sparticle
 spectrum. In addition, if SUSY is discovered at the LHC, the inclusion of 
 higher order corrections will be important for testing various SUSY breaking
 hypotheses and measuring the SUSY breaking parameters. 

\section*{Acknowledgments}
This work has been partially supported by STFC. R. RdA, is supported by the
Ram\'on y Cajal program of the Spanish MICINN and also thanks the support of
the Spanish MICINN's Consolider-Ingenio 2010 Programme under the grant
MULTIDARK CSD2209-00064, the Invisibles European ITN project
(FP7-PEOPLE-2011-ITN, PITN-GA-2011-289442-INVISIBLES and the ``SOM Sabor y
origen de la Materia" (FPA2011-29678) and the ``Fenomenologia y Cosmologia de
la Fisica mas alla del Modelo Estandar e lmplicaciones Experimentales en la
era del LHC" (FPA2010-17747) MEC projects. BCA thanks the Cambridge SUSY
working group for useful discussions.
AVB is immensely grateful to A.~Sheplyakov for providing 
a code for dealing with {\tt GiNaC} archives. The work of AVB is supported
by the RFBR grants~12-12-02-00412-a,~14-02-00494-a, and the Russian President
Grant~MK-1001.2014.2. We thank P Slavich for helpful communication and
suggestions. 

\appendix

\section{Installation of the Increased Accuracy Mode}
\label{sec:install}

The two freely available programs {\tt CLN} (at least version 1.3.1) and {\tt
  GiNaC} (at least version 1.3.5) should be installed before the user attempts
to install {\tt SOFTSUSY} with the higher order threshold corrections. However, 
{\tt SOFTSUSY} should compile without problems without these libraries if our
higher order corrections are not required. 
Two compilation arguments to the {\tt ./configure}~command are provided:
\begin{itemize}
	\item[] \verb|--enable-three-loop-rge-compilation| - compile three-loop RGEs in the MSSM\footnote{ {\tt GiNaC} and {\tt CLN} are not required 
		for three-loop RGEs.}
	\item[] \verb|--enable-full-susy-threshold-compilation| - compile
          additional two-loop threshold corrections to the third generation
          Yukawa couplings and the strong coupling constant.
\end{itemize}
Thus, if all higher order corrections are desired, {\tt CLN} and {\tt GiNaC}
should be first installed, then 
the program should be
compiled, via:
\begin{verbatim}
> ./configure --enable-three-loop-rge-compilation --enable-full-susy-threshold-compilation
> make
\end{verbatim}

We have included two global boolean variables that control the higher order
corrections  at run time, provided the program has already been compiled with 
the higher order corrections included:
\begin{itemize}
	\item \verb|USE_THREE_LOOP_RGE|  - add three-loop contribution to
          MSSM RGE (corresponds to the \code{SOFTSUSY Block} parameter 19). 
	\item \verb|USE_TWO_LOOP_THRESHOLD| - add two-loop threshold
          corrections to the third generation Yukawa couplings and the strong
          coupling constant 
          (corresponds to the \code{SOFTSUSY Block} parameter 20). If this
          variable is switched on, {\tt MssmSoftsusy} object constructors will 
          automatically include all higher order threshold corrections.
\end{itemize}
By default, both of these sets of higher order corrections are switched off 
(the boolean values are set to {\tt false}), unless the user sets them in
their main program, or in the input parameters (see~\ref{sec:run}). 

We also add the variable
\verb|double TWOLOOP_NUM_THRESH = 0.1| for finer control. It is 
          used in the iterative 
          algorithm to prevent lengthy re-evaluation of two-loop thresholds.
          If the relative difference between the two-loop thresholds obtained
          in the 
          current iteration and the value calculated in the previous
          iteration is less than
          \verb|TWOLOOP_NUM_THRESH|, the thresholds are not 
          re-evaluated for
          the next iteration.  
          See Ref.~\cite{Allanach:2001kg} for
          details of the standard {\tt SOFTSUSY} fixed point iteration 
          algorithm employed.
        
\section{Running \SOFTSUSY~in the Increased Accuracy Mode}  
\label{sec:run}

\SOFTSUSY~produces an executable called \code{softpoint.x}. 
One can run this executable from command line arguments, but the higher order
corrections will be, by default, switched off. One may switch all of the
higher order corrections on with 
the arguments 
\verb|--three-loop-rges --two-loop-susy-thresholds|
(provided they have
been compiled as specified above).
Thus, in order to produce the spectrum detailed in the $\Delta(\mbox{all})$
row of Table~\ref{tab:cmssm}, one may use the command with {\tt SOFTSUSY3.5.1}:
{\small\begin{verbatim}
./softpoint.x sugra --tol=1.0e-5 --m0=7240 --m12=800 --a0=-6000 --tanBeta=50 --sgnMu=1 --mt=173.2 
--alpha_s=0.1187 --mbmb=4.18 --two-loop-susy-thresholds --three-loop-rges 
\end{verbatim}\normalsize}
For the calculation
of the spectrum of single points in parameter space, one could alternatively use the
SUSY Les Houches Accord (SLHA)~\cite{Skands:2003cj} input/output
option. The user must provide a file (e.g.\ the example file included
in the \SOFTSUSY~distribution
\code{inOutFiles/lesHouchesInput}), that specifies the model dependent input
parameters. The program may then be run with
\small
\begin{verbatim}
 ./softpoint.x leshouches < inOutFiles/lesHouchesInput
\end{verbatim}
\normalsize

One can change whether the 3-loop RGE corrections are switched on with
\code{SOFTSUSY Block} parameter 19, whereas the 2-loop third family and $g_3$
threshold corrections 
are switched on with \code{SOFTSUSY Block} parameter 20 in the SLHA input file:
\begin{verbatim}
Block SOFTSUSY               # Optional SOFTSUSY-specific parameters
   19   1.000000000e+00      # Include 3-loop RGE terms (default of 0 to disable)
   20  31.000000000e+00      # Include all 2-loop thresholds (default of 0 to disable)
\end{verbatim}
A comment in the SLHA output file states which of the higher order terms is
included in the calculation, provided {\tt SOFTSUSY}~has been compiled to
include them. If only some of the additional two loop threshold corrections
are required, they can be switched with a finer control by changing the value
of the {\tt SOFTSUSY Block 20} parameter, as specified below.

The considered two-loop threshold corrections in a {\tt MssmSoftsusy} object are
controlled by an {\tt integer} parameter \verb|included_thresholds|. 
Depending upon the value of this integer, different approximations of the
various thresholds are included. For SUSY Les Houches Accord input,
\verb|included_thresholds| is fixed to the {\tt SOFTSUSY Block 20} parameter
input.
The various options are presented in Table~\ref{tab:bitset}.
\begin{table}
\begin{center}
\begin{tabular}{|c|ccccc|} \hline
 & \multicolumn{5}{c|}{Two-loop threshold correction} \\ 
Value & $m_t$ & $\alpha_s$ & Strong $m_b$ & Yukawa $m_b$ & Yukawa $m_\tau$ \\ 
\hline
0 & & & & & \\
1 & $\checkmark$ & & & & \\
2 & & $\checkmark$ & & & \\
3 & $\checkmark$ & $\checkmark$ & & & \\
4 & & & $\checkmark$ & & \\
5 & $\checkmark$ & & $\checkmark$ & & \\
6 & & $\checkmark$ & $\checkmark$ & & \\
7 & $\checkmark$ & $\checkmark$ & $\checkmark$ & & \\
8  & & & & $\checkmark$& \\
9  &$\checkmark$ & & &$\checkmark$ & \\
10 & &$\checkmark$ & &$\checkmark$ & \\
11 & $\checkmark$ &$\checkmark$ & &$\checkmark$ & \\
12 & & &$\checkmark$ &$\checkmark$ & \\
13 &$\checkmark$ & &$\checkmark$ &$\checkmark$ & \\
14 & &$\checkmark$ & $\checkmark$ &$\checkmark$ & \\
15 &$\checkmark$ &$\checkmark$ & $\checkmark$ &$\checkmark$ & \\
16 & & & & & $\checkmark$\\
17 &$\checkmark$ & & & &$\checkmark$ \\
18 & &$\checkmark$ & & &$\checkmark$ \\
19 &$\checkmark$ &$\checkmark$ & & &$\checkmark$ \\
20 & & &$\checkmark$ & & $\checkmark$\\
21 &$\checkmark$ & &$\checkmark$ & &$\checkmark$ \\
22 & &$\checkmark$ & $\checkmark$ & &$\checkmark$ \\
23 &$\checkmark$ &$\checkmark$ & $\checkmark$ & &$\checkmark$ \\
24 & & & &$\checkmark$ &$\checkmark$ \\
25 &$\checkmark$ & & &$\checkmark$ &$\checkmark$ \\
26 & &$\checkmark$ & &$\checkmark$ &$\checkmark$ \\
27 &$\checkmark$ &$\checkmark$ & &$\checkmark$ &$\checkmark$ \\
28 & & &$\checkmark$ &$\checkmark$ &$\checkmark$ \\
29 & $\checkmark$& &$\checkmark$ &$\checkmark$ &$\checkmark$ \\
30 & &$\checkmark$ &$\checkmark$ &$\checkmark$ &$\checkmark$ \\
31 &$\checkmark$ &$\checkmark$ &$\checkmark$ &$\checkmark$ &$\checkmark$ \\
\hline 
\end{tabular}
\end{center}
\caption{Options for finer control of the threshold loop corrections. 
A $\checkmark$ indicates that the correction is present, whereas the absence
of a $\checkmark$ indicates that the correction is absent. 
The column labelled `$m_t$' refers
to the $\mathcal O(\alpha_s^2)$ corrections to $m_t$, 
`$\alpha_s$' refers to the $\mathcal O(\alpha_s^2)$, $\mathcal O(\alpha_s \alpha_{t,b})$
  corrections to $\alpha_s$, 
`Strong $m_b$' refers to the $\mathcal O(\alpha_s^2)$ correction to $m_b$, 
`Yukawa $m_b$' refers to the $\mathcal O(\alpha_s \alpha_{t,b})$,$\mathcal O(\alpha_{b,t}^2)$, $\mathcal O(\alpha_b \alpha_t)$,
and $\mathcal O(\alpha_{t,b} \alpha_\tau)$ corrections to $m_b$ and
`Yukawa $m_\tau$' refers to the $\mathcal O(\alpha_\tau^2)$, and $\mathcal O(\alpha_\tau
\alpha_{t,b})$ corrections to $m_\tau$.
\label{tab:bitset}}
\end{table}
For convenience, we have included three {\tt MssmSoftsusy} methods that can be
used from within main programs to switch on and off some sub-classes of
threshold corrections. Each takes a {\tt bool} argument, which will switch the
correction on if it is {\tt true} and switch it off if {\tt
  false}. Table~\ref{tab:methods} displays these. 
\begin{table}
\begin{center}
\begin{tabular}{|lc|}\hline
Method & Corrections \\ \hline
{\tt setAllTwoLoopThresholds} & $\Delta m_t$, $\Delta \alpha_s$, $\Delta
m_{b},m_\tau$ \\
{\tt setTwoLoopAlphasThresholds} & $\Delta \alpha_s$ \\
{\tt setTwoLoopMtThresholds} & $\Delta m_t$ \\
{\tt setTwoLoopMbMtauThresholds} & $\Delta m_b, m_\tau$ \\
\hline\end{tabular}
\end{center}
\caption{Methods for switching on and off sub-classes of higher order
  threshold corrections. \label{tab:methods}}
\end{table}

\bibliography{threeLoop}
\bibliographystyle{elsarticle-num}
\end{document}